\documentclass[prl,aps,twocolumn,nofootinbib,floatfix,,longbibliography,superscriptaddress,notitlepage]{revtex4-2}
\usepackage[english]{babel}
\usepackage{breakcites}
\usepackage{listings}
\usepackage[utf8]{inputenc}
\usepackage[T1]{fontenc}
\usepackage{textcomp}
\usepackage{gensymb}
\usepackage{tikz}
\usetikzlibrary{positioning}

\usepackage{graphicx}
\usepackage{braket}
\usepackage{color}
\usepackage{epstopdf}
\usepackage{mathtools}
\usepackage{amsmath}
\usepackage{amssymb}
\usepackage{float}
\usepackage{comment}
\usepackage{amsmath}
\usepackage{physics}

\usepackage{amsfonts}
\usepackage{bm}
\usepackage{bbold}

\usepackage{color}

\graphicspath{{./figs/}}
\usepackage{gensymb}
\usepackage[colorinlistoftodos,prependcaption]{todonotes}
\usepackage{xargs}  
\newcommandx{\unsure}[2][1=]{\todo[linecolor=red,backgroundcolor=red!25,bordercolor=red,#1]{#2}}
\newcommandx{\change}[2][1=]{\todo[linecolor=blue,backgroundcolor=blue!25,bordercolor=blue,#1]{#2}}
\newcommandx{\info}[2][1=]{\todo[linecolor=OliveGreen,backgroundcolor=OliveGreen!25,bordercolor=OliveGreen,#1]{#2}}
\newcommandx{\improvement}[2][1=]{\todo[linecolor=Plum,backgroundcolor=Plum!25,bordercolor=Plum,#1]{#2}}
\newcommandx{\thiswillnotshow}[2][1=]{\todo[disable,#1]{#2}}
\newcommandx{\greencom}[2][1=]
{\todo[inline, color=green!40,#1]{#2}}
\newcommandx{\bluecom}[2][1=]
{\todo[inline, color=blue!40,#1]{#2}}

\usepackage[colorlinks]{hyperref}
\definecolor{winered}{rgb}{0.5,0,0}
\hypersetup
{
colorlinks=true,
linkcolor=winered,
urlcolor={winered},
filecolor={winered},
citecolor={winered},
allcolors={winered}
}

\usepackage{blindtext}
\usepackage{setspace}
\usepackage[margin=0.6in]{geometry}

\usepackage{letltxmacro}
\LetLtxMacro{\ORIGselectlanguage}{\selectlanguage}
\makeatletter
\DeclareRobustCommand{\selectlanguage}[1]{%
  \@ifundefined{alias@\string#1}
    {\ORIGselectlanguage{#1}}
    {\begingroup\edef\x{\endgroup
       \noexpand\ORIGselectlanguage{\@nameuse{alias@#1}}}\x}%
}
\newcommand{\definelanguagealias}[2]{%
  \@namedef{alias@#1}{#2}%
}
\makeatother

\definelanguagealias{en}{english}
\definelanguagealias{EN}{english}

\graphicspath{{figure/}}

\usepackage{graphicx}

\usepackage[caption=false]{subfig}

\DeclareCaptionFont{myfont}{\thecaptionfont\scriptsize\selectfont} 
\makeatletter
\renewcommand*{\fnum@figure}{{\normalfont\bfseries \figurename~\thefigure}}
\renewcommand*{\@caption@fignum@sep}{\textbf{ : }}
\makeatother

\begin{document}

\title{Gain-compensated  metal cavity modes and a million-fold improvement of Purcell factors}
\author{Becca VanDrunen}
\affiliation{\hspace{0pt}Department of Physics, Engineering Physics, and Astronomy, Queen's University, Kingston, Ontario K7L 3N6, Canada\hspace{0pt}}
\author{Juanjuan Ren}
\affiliation{\hspace{0pt}Department of Physics, Engineering Physics, and Astronomy, Queen's University, Kingston, Ontario K7L 3N6, Canada\hspace{0pt}}
\author{
Sebastian Franke}
\affiliation{Technische Universit\"at Berlin, Institut f\"ur Theoretische Physik,
Nichtlineare Optik und Quantenelektronik, Hardenbergstra{\ss}e 36, 10623 Berlin, Germany}
\affiliation{\hspace{0pt}Department of Physics, Engineering Physics, and Astronomy, Queen's University, Kingston, Ontario K7L 3N6, Canada\hspace{0pt}}
 \author{Stephen Hughes}
\affiliation{\hspace{0pt}Department of Physics, Engineering Physics, and Astronomy, Queen's University, Kingston, Ontario K7L 3N6, Canada\hspace{0pt}}


\begin{abstract}
Using a rigorous 
mode  theory for gain-compensated plasmonic dimers, we demonstrate how  quality factors and  Purcell factors can be dramatically increased, improving the quality factors from 10 to over 26,000
and the peak Purcell factors from around 3000 to over 10 billion.
Full three-dimensional calculations are presented for gold dimers in a finite-size gain medium, which allows one to easily surpass 
fundamental Purcell factor limits
of  lossy media. 
Within a regime of linear system response, we show
how the Purcell factors are modified
from the contributions from the projected local density of states as well as a non-local gain.
Further, we show that the
effective mode volume and radiative beta factors
remain relatively constant,
despite the significant enhancement of the
Purcell factors.
\end{abstract}
\maketitle 

{\it Introduction.}---Plasmonic resonators, formed by metal nanoparticles (MNPs), have become a prominent topic in nanophotonics, due in part to their unique abilities for enhancing light-matter interactions in extremely small spatial scales~\cite{Gonçalves_2020,krasnok_active_2020}. 
Plasmonic resonators exploit surface plasmons, which occur at the interface of a metal and a dielectric material, and are the result of mixed electronic-optical excitations on the surface of the metal~\cite{Pitarke_2007}. Plasmonic resonators 
yield
electromagnetic modes that are
well below the 
diffraction limit~\cite{Wang:09, Gonçalves_2020}, 
leading to improved  sensing and fast generation of single photons~\cite{Xiao_2010,8db22ac8bfcc40619fe93ded17fad762}. 

Metal based cavity modes have been explored theoretically~\cite{bai_efficient_2013-1, Hughes_SPS_2019} and experimentally~\cite{Chae_2016, RePEc:nat:nature:v:535:y:2016:i:7610:d:10.1038_nature17974}, 
for a wide range of photonics applications.
However, plasmonic resonators have significant decay rates, 
since metal is inherently a lossy material. Thus, an important goal in optical plasmonics is to find methods for alleviating this loss. 
Historically, optical mode theories of plasmonics were thought to be problematic \cite{Koenderink2010}, but this is largely caused by the use of ill-defined mode models, treating such systems like regular ``normal modes'', without much consideration for losses.

Recently, accurate cavity mode theories used to describe the optical response of MNPs have been formulated, which are based on ``quasinormal modes'' (QNMs)---the formal solution for open cavity modes~\cite{kristensen_modes_2014,kristensen_modeling_2020, ren_quasinormal_2021,lalanne_light_2018}. Similar to {\it normal modes}, QNMs are  
solutions to the source-free 
Helmholtz equation, but with \textit{open} boundary conditions, where the solutions have complex eigenfrequencies and spatially diverging fields (due to temporal losses)~\cite{kristensen_modes_2014,kristensen_modeling_2020}. 
By exploiting  QNM theory, it has become clear that cavity physics applies to plasmonic resonators~\cite{perrin_QNM_2016,lalanne_light_2018}.
Theories based on QNMs offer a significant advantage over 
all-numerical approaches, which can be tedious, limited in scope, and often do not even explain the basic mechanisms of field enhancement. 
In contrast, a 
mode theory provides many physical insights, 
is efficient, has wide applications, and
lends itself 
to mode quantization~\cite{franke_quantization_2019}.
Moreover, the modes form a basis for computing the
photonic Green function (GF), which can be used to describe
a wide range of optical phenomena in both classical and quantum optics~\cite{vlack_dyadic_2012,ge_quasinormal_2014,dung_three-dimensional_1998,ren_quasinormal_2021, Righini_2007_trapping}. 

Accounting for both material loss and radiative loss, the mode quality factor is defined from $Q=\omega_c/(2\gamma_c)$ ($2\gamma_c$ is the energy loss rate of the mode {\it c}), which for plasmonic resonators is much smaller than typical dielectric cavities~\cite{sharma_2018_dielectric}. Thus,  the MNP resonances  typically generate quality factors of around $Q \approx 10-20$~\cite{cognee_cooperative_2019, melli_2013_Qlim, Lilley:15,kuttge_nanodisk_2010},
which manifest in a very short cavity mode lifetime, $\tau_c = {1}/{\gamma_c}$, 
e.g., a resonance of  
$\hbar \omega_c = 1.780 - i0.068$~eV~\cite{ren_near-field_2020}, corresponds to  $\tau_c \approx 
0.01$~ps. Such losses prohibit many applications in coherent optics, including 
surface plasmon lasing  and spasing~\cite{stockman_spasers_2008, stockman_spaser_2011}.

One potential method for mitigating this 
significant loss is through gain compensation, which uses material gain to suppress some of the dissipation, using  linear amplification~\cite{stockman_gain_2011,russev_conditions_2012, veltri_PRB_2012}.
It is important to note that the total cavity structure must be overall lossy for the properties of cavity physics and  QNM theory to apply, and also to maintain a  {\it linear medium} response. More specifically, the entire GF is only allowed to have complex poles in the lower complex half plane.
To model such a structure, both the metal and the gain 
are  defined by a \textit{complex} dielectric constant, where the imaginary 
part describes loss or gain.

Gain media has been utilized in several applications to suppress metallic losses~\cite{De_Leon_Berini_2010,Noginov_2008,Berini_De_Leon_2011}. For example, loss suppression in MNPs has been studied by doping them in gain~\cite{Pustovit:15,liu_efficient_2011} or by adding gain to plasmonic resonators~\cite{Cai:18,Fang:11}.
While there 
exists some studies on how the quality factor changes as a result of gain-compensation~\cite{liu_efficient_2011,Cai:18,Ding_2013,Wang_nanolaser_2017},
little has been done to quantify how enhanced spontaneous emission (SE) and Purcell factors change from coupled dipole emitters.
Furthermore, the concept of spasing has been an enticing area of study for decades~\cite{Zhong_spaser_2013,Zhong_Hong_Li_2013,Warnakula_2018}, yet many approaches lack the robustness of a rigorous mode theory.

An important metric
in nanophotonics
is the Purcell factor~\cite{gaponenko_2010},
which describes
 the enhanced SE rate of a dipole emitter, $\Gamma(\mathbf{r}_0,\omega)$,
 normalized to the rate from a background homogeneous medium, $\Gamma_{\rm B}(\mathbf{r}_0,\omega)$, which is
   $ F_{\rm P} \equiv \frac{3}{4 \pi^2}\left(\frac{\lambda_0}{{n}_{\rm B}}\right)^3\frac{Q}{V_{\rm eff}}$,
where $Q$ and $V_{\rm eff}$ are the mode quality factor and effective mode volume, respectively, $\lambda_0$ is the free space wavelength, and  $n_{\rm B}$ is the background refractive index. This well known formula
assumes the emitter is on resonance with the cavity mode and at a field maximum position, with the same polarization as the cavity mode. 
 Traditionally, the effective mode volume describes the volume associated with mode localization, but that is no longer correct using QNMs
 , 
 and the generalized mode volume can be both complex and position dependent \cite{kristensen_calculation_2014}. 
 Moreover, 
when gain is added to the cavity,
this formula no longer applies (even when using QNMs), and one 
must include a non-local correction from gain~\cite{franke_fermis_2021,ren_classical_2023}. 

In this work, we
show how material gain can significantly improve the MNP cavity mode properties,
using a rigorous QNM theory with linear amplification. The key findings are:
(i)  the effective mode volume of the mode profiles stays nearly constant as gain is introduced to the system;
(ii) the Purcell factor yields a million-fold improvement when material gain is added; and (iii) the
radiative beta factor also remains relatively constant. Our gain theory has a wide range of applications, including quantum sensing, as well as lasing
and fundamental topics in quantum topics.

{\it Theory.}---The QNMs, ${\Tilde{\bf f}}_\mu$, 
are the mode solutions 
to the Helmholtz equation, 
with open boundary conditions~\cite{ren_quasinormal_2021}:
\begin{equation}
    {\bm \nabla} \cross {\bm \nabla} \cross \Tilde{{\bf f}}_\mu({\bf r}) - \left(\frac{\Tilde{\omega}_{\mu}}{c} \right)^2 \epsilon({\bf r}, \Tilde{\omega}_\mu)\Tilde{{\bf f}}_\mu({\bf r}) = 0,
    \label{eq1}
\end{equation}
where $c$ is the speed of light in a vacuum, $\Tilde{\omega}_\mu$ is the QNM complex eigenfrequency $\Tilde{\omega}_\mu = \omega_\mu - i\gamma_\mu$, and $\epsilon({\bf r}, \Tilde{\omega}_\mu)$ is the complex dielectric function.
The inhomogeneous 
Helmholtz equation for an arbitrary polarization source  
can be used to define the
GF, from
$
c^2\mathbf{\nabla}\times\mathbf{\nabla}\times{\mathbf{G}}(\mathbf{r},\mathbf{r}^\prime,\omega) -\omega^2\epsilon(\mathbf{r},\omega){\mathbf{G}}(\mathbf{r},\mathbf{r}^\prime,\omega) \nonumber 
=  \omega^2 {\mathbf{1}}\delta(\mathbf{r} - \mathbf{r}'),
$
where 
the electric field solution is at $\mathbf{r}$, when a source field dipole is at $\mathbf{r}^\prime$.
Within or near the cavity region, 
the GF can be expressed as a sum of normalized QNMs~\cite{ren_near-field_2020, leung_completeness_1994, ge_quasinormal_2014}:
\begin{equation}
    {\bf G}_{\rm }({\bf r}, {\bf r}_0, \omega) = \sum_{\mu} A_{\mu}(\omega) \Tilde{{\bf f}}_{\mu}({\bf r}) \Tilde{{\bf f}}_{\mu}({\bf r}_0),
    \label{eq2}
\end{equation}
where $A_{\mu}(\omega)= {\omega}/{[2(\Tilde{\omega}_\mu - \omega)]}$~\cite{bai_efficient_2013-1},
and we note the vector product is unconjugated (i.e., not $\tilde {\bf f}_\mu \tilde {\bf f}_\mu^*$), which is a consequence of using a non-Hermitian theory.
When a single QNM dominates,
$\mu = c$, then the GF
is simply
\begin{equation}
    {\bf G}_c({\bf r}, {\bf r}_0, \omega) \approx A_c(\omega) \Tilde{{\bf f}}_c({\bf r}) \Tilde{{\bf f}}_c ({\bf r}_0).
    \label{eq4}
\end{equation}

In a lossy material system with no gain, the SE rate for a dipole emitter at a location ${\bf r}_0$ is determined from the (projected) local density of states (LDOS)~\cite{kristensen_modes_2014, ren_near-field_2020}:
\begin{equation}
    \Gamma_{\rm LDOS}^{\rm SE}({\bf r}_0,\omega) = \frac{2}{\hbar \epsilon_0}{\bf d} \cdot {\rm Im}[{\bf G}({\bf r}_0, {\bf r}_0, \omega)]\cdot {\bf d},
    \label{eq5}
\end{equation}
and using a single QNM expansion approximation, ${\bf G}$ becomes ${\bf G}_c$, as shown in Eq.~\eqref{eq4}. 
The SE rate for a dipole  in a homogeneous medium, $\Gamma_{\rm B}({\bf r}_0,\omega)$, is similarly obtained by replacing 
${\bf G}$ by 
$\mathbf{G}_{\rm B}$ (the GF of a homogeneous medium), which is known analytically.
The LDOS Purcell factor is then given by \cite{Anger2006, kristensen_modes_2014, ren_near-field_2020}
%
\begin{align}
    F_{\rm P} ^{\rm LDOS}({\bf r}_0, \omega) 
    &
    = 1 + \frac{6 \pi c^3}{\omega ^3 n_{\rm B}}{\bf n}_{\rm d} \cdot {\rm Im}[{\bf G}_{} ({\bf r}_0,{\bf r}_0, \omega)] \cdot {\bf n}_{\rm d}
    \label{eq: LDOS PF},
\end{align}
and $F_{\rm P}^{\rm QNM, LDOS}$
is obtained by 
using ${\rm G} \rightarrow {\rm G}_c$.

However, this classical LDOS formalism for the Purcell factor is not correct in the presence of a linear gain medium.
Instead, the total SE rate can be written as~\cite{franke_fermis_2021,ren_classical_2023} 
\begin{align}\label{eq: Gamma_tot}
\Gamma^{\rm SE}_{\rm tot}({\bf r}_0,\omega) = \Gamma^{\rm SE}_{\rm LDOS}({\bf r}_0,\omega) + \Gamma^{\rm SE}_{\rm gain}({\bf r}_0,\omega),
\end{align}
which notably contains 
an extra net-positive term related to gain region added to the traditional LDOS term; this correction can be derived quantum mechanically~\cite{franke_fermis_2021,ren_quasinormal_2021} or classically~\cite{ren_classical_2023}.
The total Purcell factor is then
\begin{equation}
 \label{eq: total PF}
 F_{\rm P}(\mathbf{r}_{0},\omega) =  
 1 +  \frac{\Gamma^{\rm SE}_{\rm tot}(\mathbf{r}_{0},\omega)}{\Gamma_{\rm B}(\mathbf{r}_{0},\omega)}.
 \end{equation}

In origin, 
the well known LDOS-SE 
formula 
is linked to the GF identity, $\int_{\mathbb{R}^3}{\rm d}\mathbf{s}\epsilon_I(\mathbf{s})\mathbf{G}(\mathbf{r},\mathbf{s})\cdot\mathbf{G}^*(\mathbf{s},\mathbf{r}')={\rm Im}[\mathbf{G}(\mathbf{r},\mathbf{r}')]$, which involves an integration over all space. However, from this identity, one must subtract the contribution from the gain, and this results in \textit{adding} the separate gain contribution term, 
which is given by~\cite{franke_fermis_2021}
\begin{equation}
    \Gamma_{\rm gain}^{\rm SE}(\mathbf{r}_0,\omega) = \frac{2}{\hbar \epsilon_0}{\bf d} \cdot {\bf K}({\bf r}_0, {\bf r}_0, \omega) \cdot {\bf d},
    \label{eq: quan gamma}
\end{equation}
where
\begin{equation}
\mathbf{K}(\mathbf{r}_0,\mathbf{r}_0,\omega)
=\!\int_{V_{\rm gain}}\!d\mathbf{s}\left|{\rm Im}[\epsilon^{\rm gain}(\mathbf{s})]\right|\mathbf{G}(\mathbf{r}_0,\mathbf{s},\omega)\cdot\mathbf{G}^{*}(\mathbf{s},\mathbf{r}_0,\omega).
\end{equation}
In the case of a single dominant QNM, 
 $\mu = c$, then
\begin{align}
\begin{split}
\Gamma_{\rm gain}^{\rm SE}({\bf r}_0, \omega)
&=\frac{2|\mathbf{d}|^2}{\hbar \epsilon_0}\Big|A_{c}(\omega)\Big|^2\Big|\mathbf{n}_{\rm d}\cdot \Tilde{{\bf f}_c}(\mathbf{r}_{0})\Big|^2 \times \\
&\int_{V_{\rm gain}}d\mathbf{s}\Big|{\rm Im}[\epsilon^{\rm gain}(\mathbf{s})]\Big|\Big|\Tilde{{\bf f}_c}({\bf s})\Big|^2.
\label{eq:gain-cav}
\end{split}
\end{align}

From the perspective of 
classical power flow arguments~\cite{ren_classical_2023}, the total Purcell factor can be written as
\begin{align}
\label{eq: FP_mod2}
    F_{\rm P}^{\rm }(\mathbf{r}_0,\omega)=
     \frac{P_{\rm LDOS}(\mathbf{r}_0,\omega) + P_{\rm gain}(\mathbf{r}_0,\omega)}{P_0({\omega})},  
\end{align}
where $P_{\rm LDOS}$ and $P_0$ are the power flow from the point dipole with and without the cavity structure (background medium), respectively, and   $P_{\rm gain}$ is the power flowing {\rm out} from the gain region (net-positive).
Equation~\eqref{eq: FP_mod2}
can be solved using QNMs or numerically.
%
The total Purcell factors can also be written as 
$F_{\rm P}^{\rm }(\mathbf{r}_0,\omega)=
     \frac{P_{\rm far}(\mathbf{r}_0,\omega) + P_{\rm loss}(\mathbf{r}_0,\omega)}{P_0({\omega})}$,  
where $P_{\rm far}$ and $P_{\rm loss}$ are power radiated to far field region and dissipated within lossy region, respectively. 

From this general SE decay theory, one can also determine the radiative beta factor ($\beta$-factor), which represents the probability
that an emitted photon will
decay radiatively to the far field~\cite{ren_near-field_2020}. 
Generally, there are both radiative and non-radiative $\beta$-factors, and the sum of these must equal one~\cite{Hughes_SPS_2019}. The radiative beta factor is~\cite{ren_classical_2023}
\begin{equation}
    \beta^{\rm rad}(\mathbf{r}_0, \omega) = \frac{P_{\rm far}(\mathbf{r}_0, \omega)}{P_{\rm far}(\mathbf{r}_0, \omega) + P_{\rm loss}(\mathbf{r}_0, \omega)},
    \label{rad beta factor}
\end{equation}
which can also be written equivalently in terms of $P_{\rm LDOS}$ and $P_{\rm gain}$ based on the power conservation law~\cite{ren_classical_2023}.

\begin{figure}[t]
    \begin{tikzpicture}
    \centering
        [inner sep=0mm]
 \node[xshift=-10mm] (figa) 
{\includegraphics[width=0.40\columnwidth]{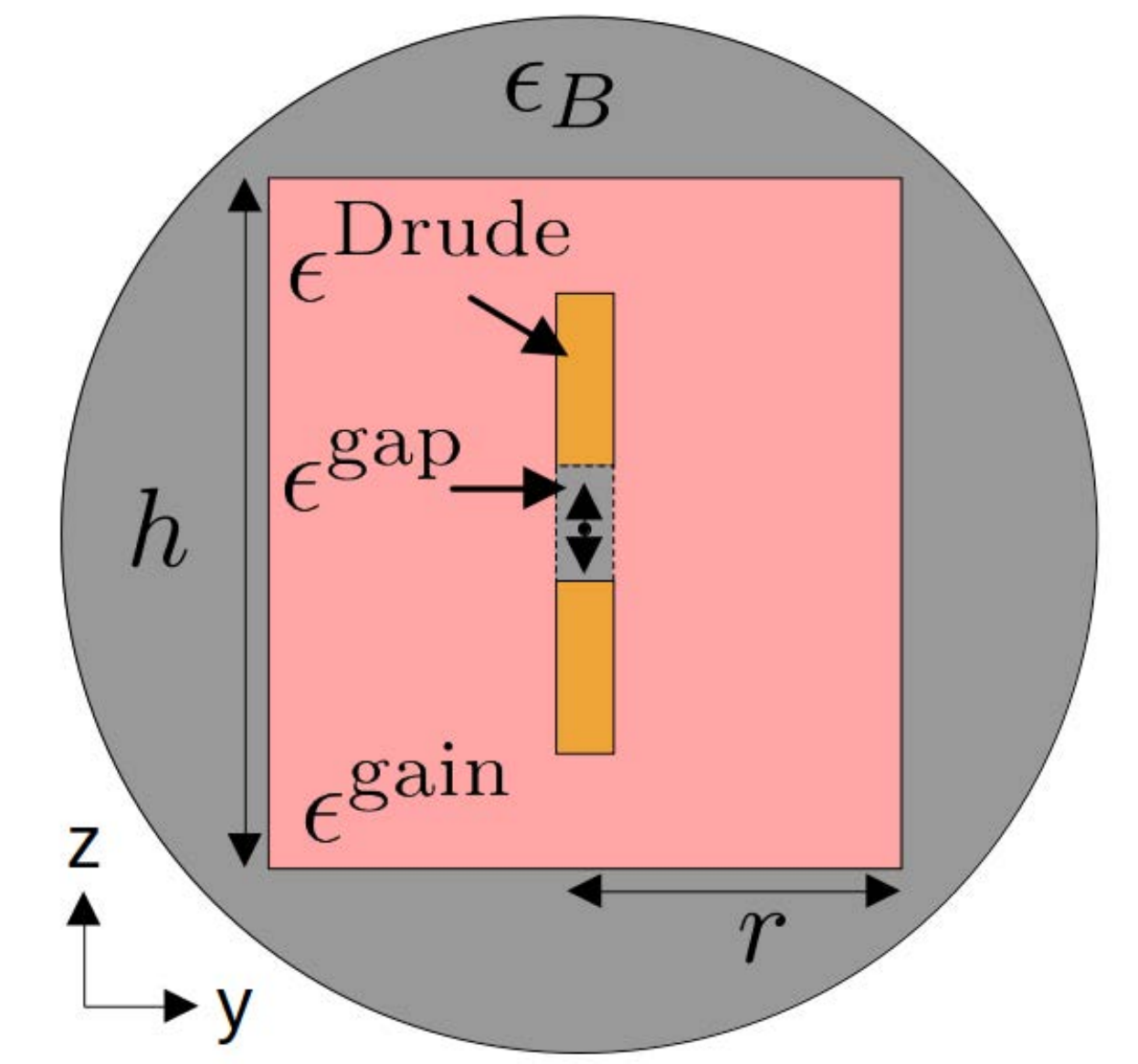}};
 \node[right=of figa,xshift=-10mm] (figb) {\includegraphics[width=0.35\columnwidth]{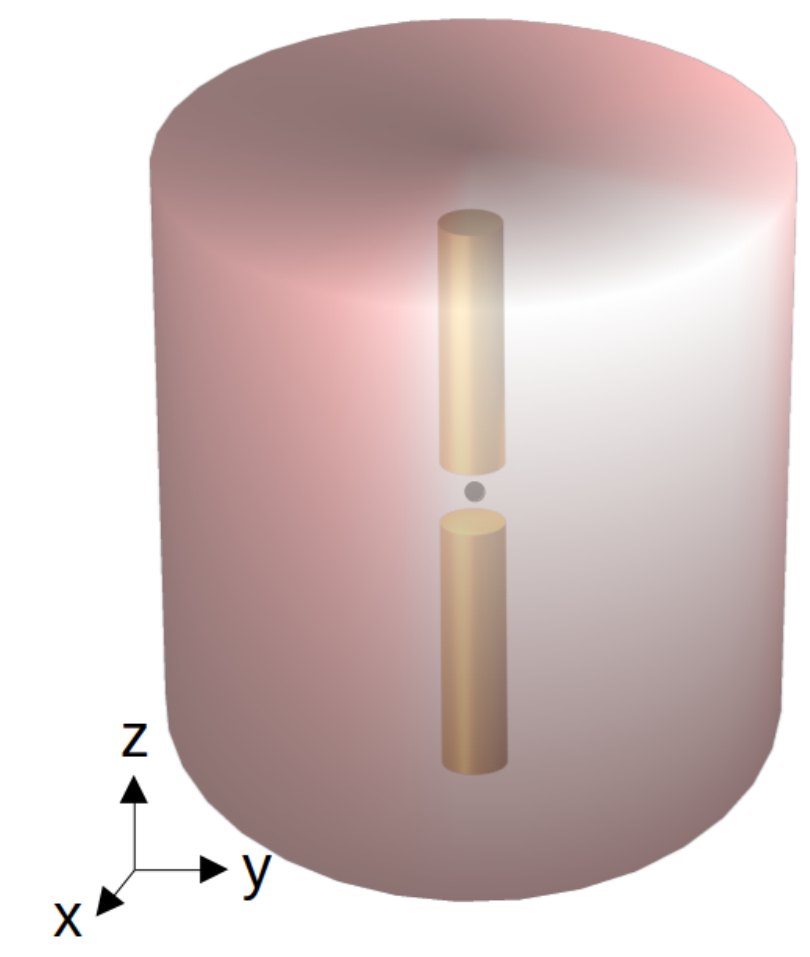}};
\node[xshift=-31mm, yshift=4mm] at (figa.north east) {(a)};
 \node[xshift=16mm, yshift = 2mm] at (figb.north west) {(b)};
\end{tikzpicture}
\vspace{-0.5cm}
\caption{(a) A  2D view of the resonator system. The dielectric functions for each material are labelled, where the background medium has $\epsilon_{\rm B} = 2.25$ ($\epsilon_{\rm B} = n_{\rm B}^2$), the gain region has $\epsilon^{\rm gain} = 2.25 - i\alpha_g$ (where $\alpha_g$ is the gain parameter), the gold nanorods have $\epsilon^{\rm Drude}$ which is governed by the Drude model (see text), and the gap region has $\epsilon^{\rm gap}=2.25$. The gold nanorods have a length of $80~$nm, a radius of $10~$nm, and the gap distance is $20~$nm; $h$ represents the height of the gain region, and is $400~$nm, and $r$ represents the radius of the gain region, which is $200~$nm. (b)  3D version of the system (as used in our model).
}
\label{fig: model}
\end{figure}

{\it Results.}---The main MNP cavity structure we model consists of a gold dimer  enclosed in a finite-sized cylindrical region of gain, as shown in Fig.~\ref{fig: model}. 
The dielectric constant for the gold is  described by the Drude model,
  $  \epsilon^{\rm Drude}(\omega)=1-\frac{\omega_{\rm p}^{2}}{\omega^{2}+i\omega\gamma_{\rm p}}$,
with $\hbar\omega_{\rm p}=8.2934$ eV 
and $\hbar\gamma_{\rm p}=0.0928$ eV. 
The gap region between the nanorods (where the dipole sits) is considered to have a real permittivity. The gain region has a permittivity of $\epsilon^{\rm gain} = 2.25 - i\alpha_g$, so the gap region simply takes the real component of this. A real example of a gain material is Rh6G dye in PMMA, e.g., with permittivity of $\epsilon^{\rm gain} = 2.25 - i0.006$~\cite{russev_conditions_2012,Noginov:08}. Gallium Arsenide quantum wells can also have rather large gain values, e.g., $\epsilon^{\rm gain} = 11.76 - i0.208$~\cite{russev_conditions_2012,CORZINE199317}. 
We assume the gain is dispersionless, but 
gain dispersion results are discussed
in the Supplementary Material (SM)~\cite{SM}, and do not affect our general findings.
For all calculations below, 
we  obtain the
QNMs numerically using a complex frequency approach, implemented in COMSOL~\cite{bai_efficient_2013-1,ren_near-field_2020}.

\begin{figure}[h!]
    \centering\includegraphics[width=0.82\columnwidth]{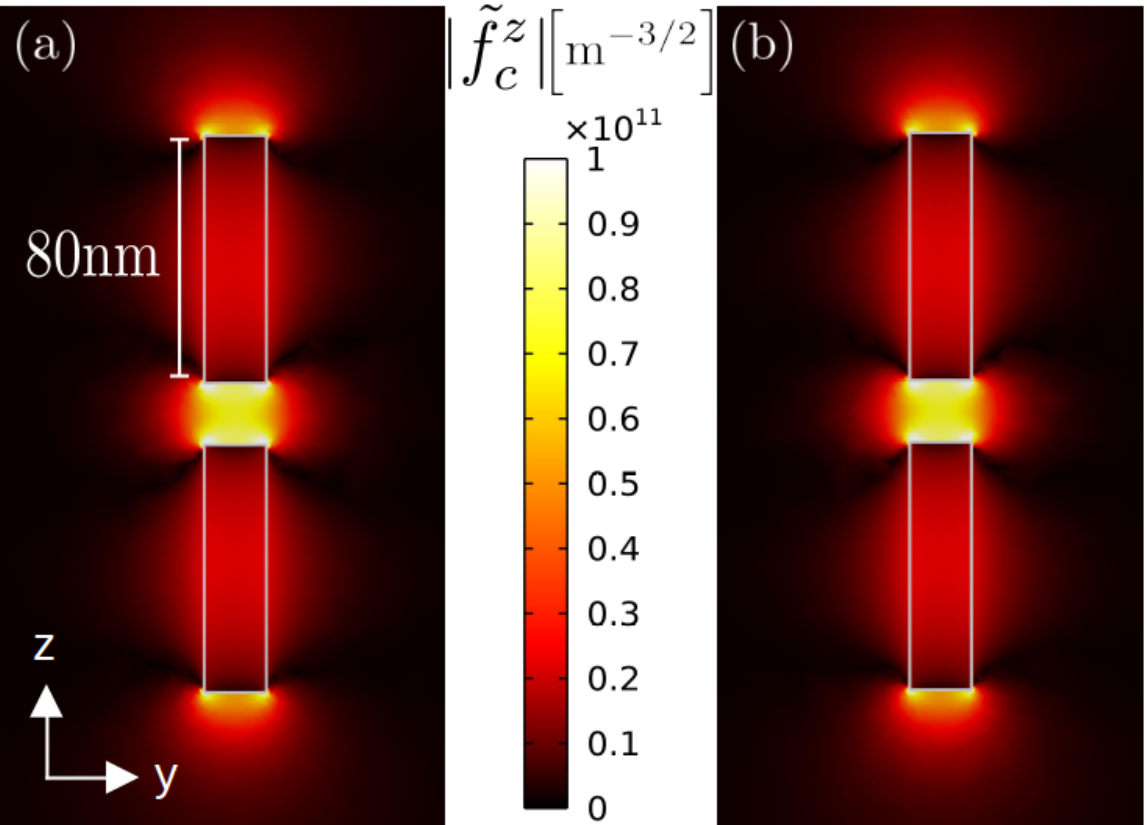}  
    \caption{ Surface plots  of the computed 
   QNM profile (using the dominant field component) for the plasmonic resonator system (a) without  gain, and (b) with gain, using  $\alpha_g = 2.54 \cdot 10^{-1}$.
    }
    \label{fig:mode volumes}
\end{figure}

\begin{figure*}
\subfloat[small gain values]{\includegraphics[width=0.59\columnwidth]{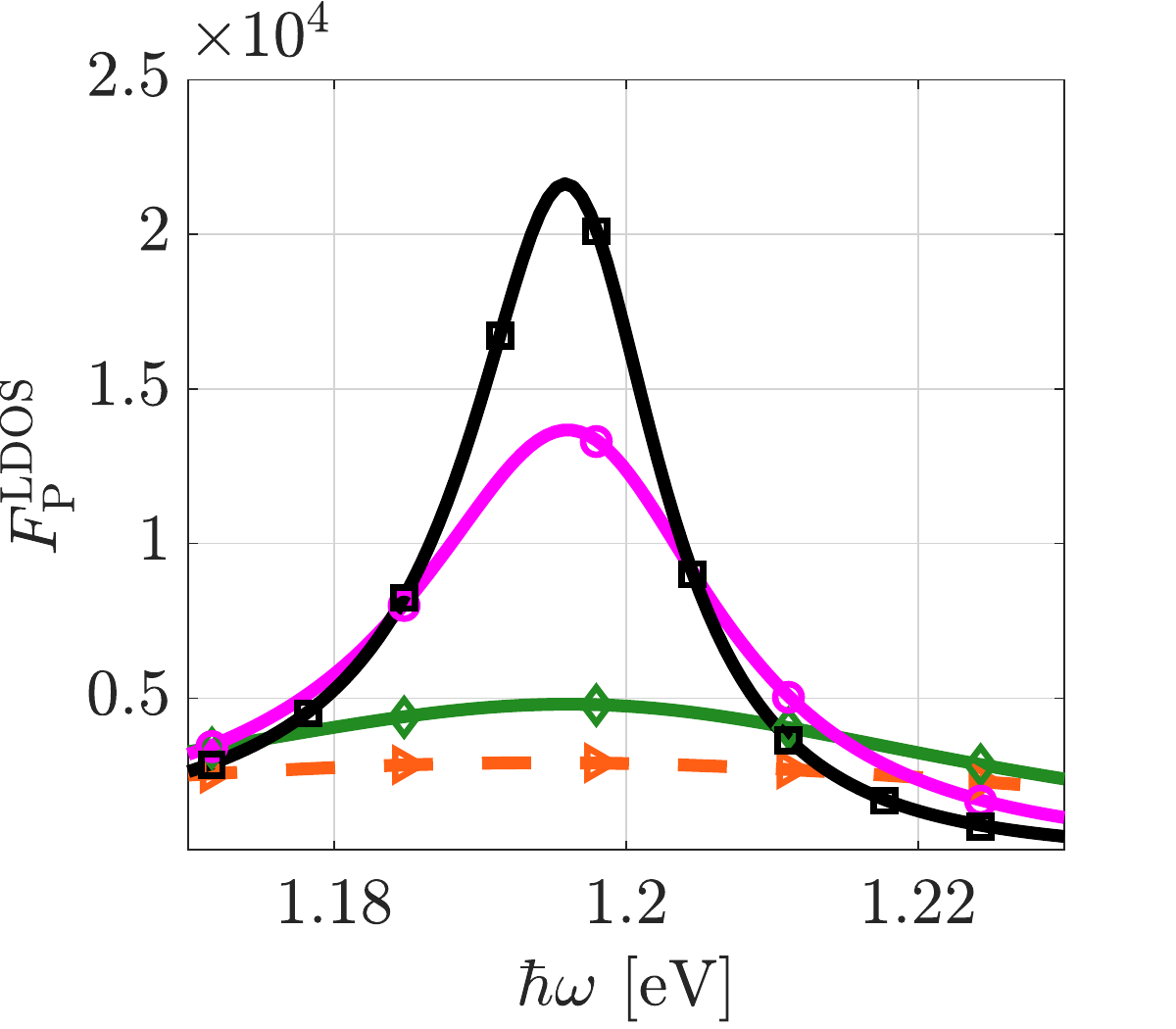}\label{pf_small_gain}}
\subfloat[larger gain values]{\includegraphics[width=0.59\columnwidth]{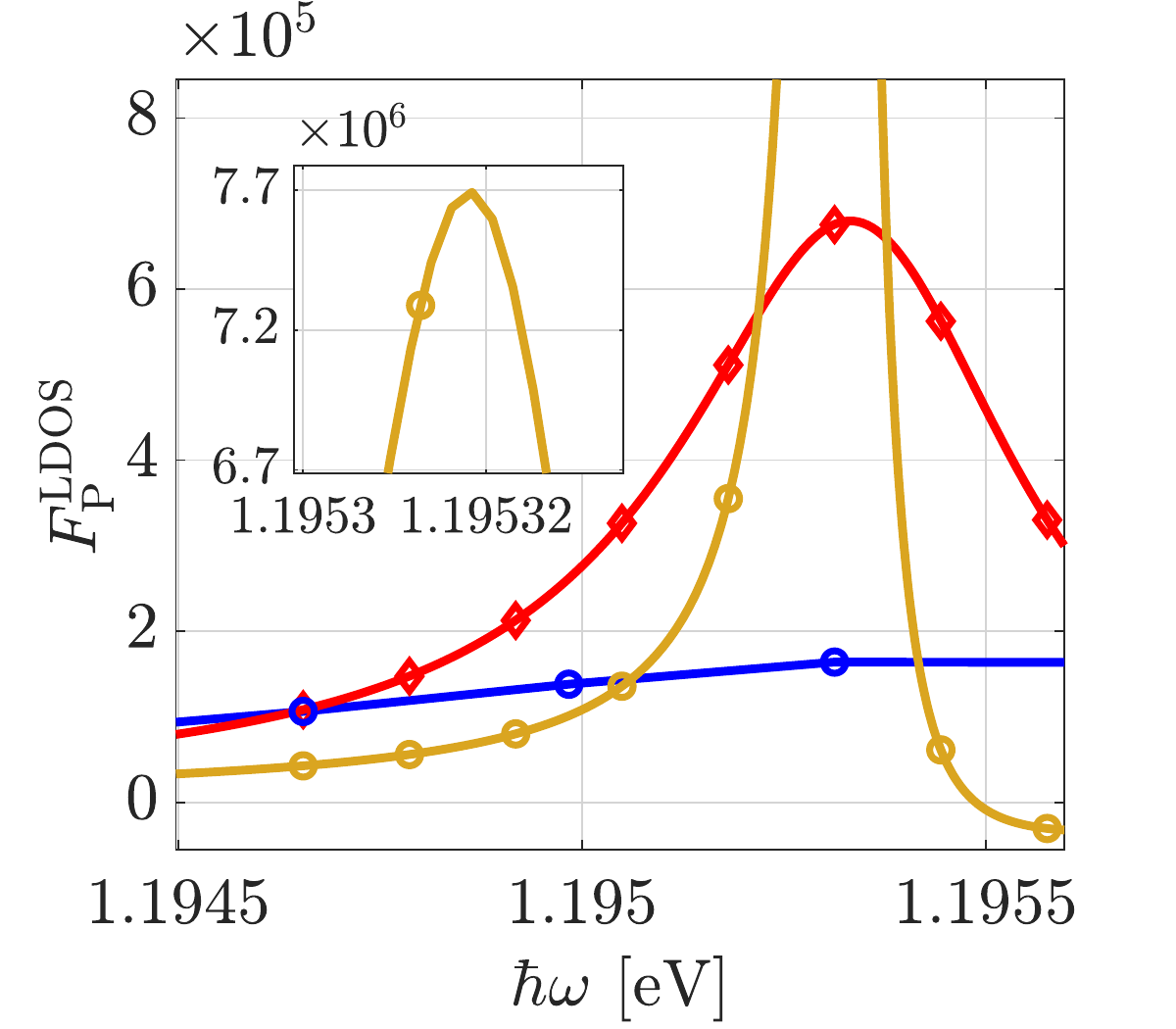}\label{pf_big_gain}}
\subfloat[total Purcell factors, for cases in (b)]{\includegraphics[width=0.59\columnwidth]{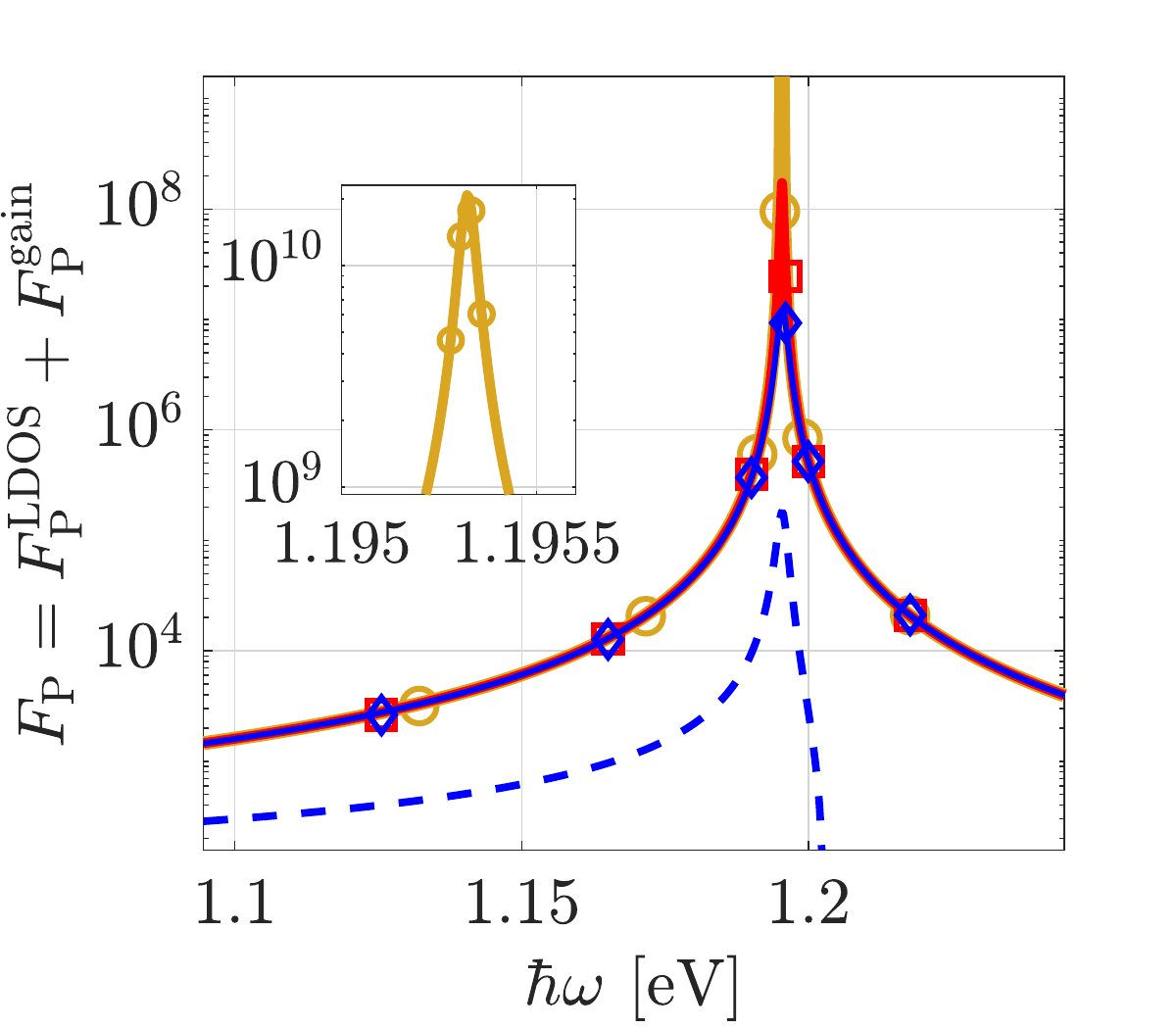}\label{pf_quan}}
\caption{
(a) LDOS Purcell factors for various $\alpha_g$, 
corresponding to the model in Fig.~\ref{fig: model}. The curves show the QNM method, calculated with Eq.~\eqref{eq: LDOS PF}, and the symbols show the full dipole numerical solution with Eq.~\eqref{eq: FP_mod2}. The orange dashed line/symbols are for the case with  $\alpha_g = 0$, the green line/points represent $\alpha_g = 1\cdot 10^{-1}$, the magenta line/points represent $\alpha_g = 2\cdot 10^{-1}$, and the black line/points represent $\alpha_g = 2.2\cdot 10^{-1}$.
The $\gamma_c$ values for each curve in increasing order of $\alpha_g$ are: $\gamma_0$, $0.61 \gamma_0$, $0.21 \gamma_0$, $0.13 \gamma_0$
, where $\gamma_0 = 9.016 \cdot 10^{13}$ rads/s. 
(b) LDOS Purcell factors for larger values of $\alpha_g$.
The blue line/points represent the case where $\alpha_g = 2.5\cdot 10^{-1}$, the red line/points represents the case where $\alpha_g = 2.53\cdot 10^{-1}$, and the gold line/points represents the case where $\alpha_g = 2.54\cdot 10^{-1}$.
The $\gamma_c$ values for each curve in increasing order of $\alpha_g$ are: $0.02 \gamma_0$, $0.004 \gamma_0$, and $0.0004 \gamma_0$. The bandwidth ($2\gamma_c$) for each curve in increasing order of $\alpha_g$ are: $2.4~$meV, $0.48~$meV, and $0.048~$meV.
(c) 
\textit{Total} Purcell factor over a range of energies for three different values of $\alpha_g$, calculated with Eq.~\eqref{eq: total PF} for the QNM method, and a full numerical dipole method to confirm the results, using Eq.~\eqref{eq: FP_mod2}. The blue, red, and gold line/points correspond to the same $\alpha_g$ values used in Fig.~\ref{pf_big_gain}, all plotted on a logarithmic $y$-axis. The dashed blue line represents the LDOS Purcell factor when $\alpha_g = 2.5\cdot 10^{-1}$, which shows a significant difference (and becomes negative at larger frequencies). The bandwidths in increasing order of $\alpha_g$ are: $2.4~$meV, $0.48~$meV, and $0.048~$meV.
}
  \label{fig: pf with different gain}
\end{figure*}

The vector-field QNM is dominated by $z$-polarization, that peaks in frequency near
$\hbar\omega \approx 1.2~$eV. 
As can be seen in Fig.~\ref{fig:mode volumes}, the spatial profile of the QNM is similar with and without the gain medium. Furthermore, this can be verified quantitatively, by calculating the effective mode volume, $V_{\rm eff}^{-1}({\bf r}_0)=\epsilon(\mathbf{r}_{0}){\rm Re}[\tilde{\bf f}_c^2({\bf r}_0) ]$
from QNM theory. 
See SM~\cite{SM} for the  computed  effective mode volume 
at gap center. In stark contrast to lasing modes, and constant-flux modes, which can observe drastic changes to the spatial modes, the linear-gain-assisted modes remain relatively similar~\cite{PhysRevA.84.023820_Ge, PhysRevLett.129.133901_Doronin}.

For the two cases shown in Fig.~\ref{fig:mode volumes}, the complex QNM eigenfrequencies are $\hbar\Tilde{\omega}_c =1.198-i5.934\cdot 10^{-2} ~{\rm eV}$ and $\hbar\Tilde{\omega}_c = 1.195 - i2.238 \cdot 10^{-5}  ~{\rm eV}$,  respectively. This leads to QNM quality factors $Q$ of 10 and 26,698, respectively, showing that a significant improvement in $Q$ is possible.

In the presence of gain, 
the LDOS Purcell factor of a dipole placed at the dimer gap center
can be calculated using Eq.~\eqref{eq: LDOS PF} and checked against a full numerical dipole calculation~\cite{ren_near-field_2020,ren_classical_2023}. 
The LDOS Purcell factor gives insights into how the gain medium impacts the SE rates. 

A summary of the Purcell factors with and without gain is 
shown in Fig.~\ref{fig: pf with different gain}.
Fig.~\ref{pf_small_gain} shows the LDOS Purcell factors
for  $\alpha_{g}$ from
$0~{\rm to}~2.2 \cdot 10^{-1}$; 
the agreement between the QNM method (curves) and the full dipole method (symbols) is excellent, and we stress there are no fitting parameters used in this model. A summary of the Purcell factor at a fixed frequency over a range of spatial points within the gap center can be found in the SM~\cite{SM}. 
This exhibits the power of our QNM theory and also shows that significant Purcell factors occur over a large spatial domain.

Moreover, 
both Figs.~\ref{pf_small_gain} and \ref{pf_big_gain} (which increases the gain to 
$\alpha_{g}=2.5\cdot 10^{-1}\sim2.54 \cdot 10^{-1}$) demonstrate a substantial increase in the LDOS Purcell factor. 
Specifically, 
the LDOS Purcell factor experiences an increase by a factor of nearly 3000, between the LDOS Purcell factors for $\alpha_g = 0$ and $\alpha_g = 2.54\cdot 10^{-1}$.
Since $V_{\rm eff}$ is similar with and without gain, this increase is primarily caused by {\it gain compensating} the loss. As anticipated, there are values of the LDOS Purcell factor that are negative for certain frequency values (as the LDOS becomes negative).
We note again that there is nothing unphysical about a negative LDOS; it means that there is more local power flow back to the dipole location 
than out of the dipole, but the entire cavity system is still net lossy.
The sign of the LDOS Purcell factor is dependent on position, and does not relate to other quantities such as the quality factor~\cite{franke_fermis_2021}. Negative LDOS Purcell factors have no implications on linear amplification either.

Next, we show the 
 total Purcell factor in Fig.~\ref{pf_quan}, corresponding to the $\alpha_g$ values from Fig.~\ref{pf_big_gain}. There are no longer any negative values
for the total Purcell factor, and the (correct) enhanced SE has been further increased, 
by many orders of magnitude,
peaking at a value of
$2.08 \cdot 10^{10}$
when the gain value is
$2.54 \cdot 10^{-1}$, yielding an increase by a factor of $7.1\cdot 10^{6}$ from the case with no gain.
As anticipated by our analytical QNM theory, the lineshapes also deviate significantly from Lorentzian and become similar to Lorentzian-squared when the gain contribution dominates, sharing some features and applications of resonances near exceptional points~\cite{pick_general_2017,miri_exceptional_2019,ren_quasinormal_2021}, such as enhanced sensing.

As mentioned before, it is crucial for the region of {\it linear amplification} that we ensure the QNM peak eigenfrequency $\tilde{\omega}_{\rm c}=\omega_{\rm c}-i\gamma_{\rm c}$ has a positive value for $\gamma_{\rm c}$. This threshold has been found heuristically~\cite{liu_efficient_2011}, however the QNM method provides a more rigorous definition of the linear amplification regime through the resonant eigenfrequency.

It is important to note that 
the gain contribution to enhancing the LDOS Purcell factor (with gain) does not really affect the 
bandwidth,
and is precisely
$2\gamma_c$ with a lineshape
of a Lorenzian squared function
[see Eq.~\eqref{eq:gain-cav}].
This is quite different to lossy media and dielectric cavities, where the Purcell factors 
are often maximized 
to near 1000, and the $Q$ factors to around 1 million
(for example, see Refs.~\cite{10.1063/1.4991416, vasco2021global,granchi2023q}),
and the bandwidths are much narrower, since these are inversely proportional to $Q$. The narrow bandwidths
in high $Q$ dielectric cavities 
are also influenced from manufacturing disorder, so further improvements, even with complex design, are not so beneficial.
As a state-of-the-art example with sub-wavelength mode volumes, impressive 
topologically-optimized dielectric cavities can yield a peak Purcell factor on the order of $F_P = 6 \cross 10^{3}$~\cite{albrechtsen2022nanometer}, with a measured value of roughly $F_P = 1 \cross 10^{3}$, 
due to manufacturing imperfections and disorder. 

We note that for our peak Purcell values
exceeding 10 billion, the bandwidth
is around $0.048~$meV, while a 1 million quality-factor 
dielectric design working at 200 THz (1.5 microns)
has a much smaller bandwidth of $0.00074$ meV.

Finally, we examine the modified beta factors,
using Eq.~\eqref{rad beta factor}.
In the linear regime, the $\beta$-factor must have an upper limit of 1, which is the $\beta$-factor for an ideal lossless dielectric system, and values surpassing this limit are indicative of the lasing regime~\cite{PhysRevB_Chelsea}. With no gain, the beta factor from the metal dimer cavity (on resonance) is 0.33, and when $\alpha_g = 0.254$, the beta factor is 0.376.
Further insights about the beta factor can be found in the SM~\cite{SM}. 
Having a large radiative beta factor 
is useful for many applications, such as 
 lasing/spasing and sensing applications.~\cite{PhysRevLett.118.237402,mikhailova2019nanostructures,gao_nanolaser_2020,Caicedo:22}.

In summary,
using a rigorous and powerful QNM theory, the enhanced SE rates of a
dipole emitter in a plasmonic resonator system were studied, with and without gain. Using material gain to compensate for the lossy nature of the gold dimer, the LDOS and total Purcell factors were shown to be substantially increased; the Purcell factor with no gain peaks around $2900$, but with a maximum gain value of $\alpha_g = 2.54\cdot 10^{-1}$ (for linear amplification) the LDOS and total Purcell factors peak at $7.7\cdot 10^{6}$ and $2.08\cdot 10^{10}$, respectively.  
We also discussed how we ensure a regime of linear amplification, and
  demonstrated that a single QNM theory worked quantitatively well by comparing with numerically exact simulations (subject to numerical limitations).
Determining how gain-compensation of loss impacts properties, such as the enhanced SE, is important for developing accurate models of plasmonic lasers, quantum sensors, and lossy cavity systems that can possibly achieve the regime of ultrastrong light-matter coupling
($g/\omega_c>0.1$~\cite{FriskKockum2019,RevModPhys.91.025005,Salmon2022}, where $g$ is the cavity-dipole coupling rate).

We  acknowledge funding from Queen's University, Canada, 
the Canadian Foundation for Innovation (CFI), 
the Natural Sciences and Engineering Research Council of Canada (NSERC), and CMC Microsystems for the provision of COMSOL Multiphysics.
We also acknowledge support from the 
Alexander von Humboldt Foundation through a Humboldt Research Award.

\textit{Disclosures.}---The authors declare no conflicts of interest.

 \textit{Data Availability.}---Data underlying the results in this paper are not publicly available at this time but may be obtained from the authors upon reasonable request.



\bibliography{refs}

\begin{thebibliography}{69}%
\makeatletter
\providecommand \@ifxundefined [1]{%
 \@ifx{#1\undefined}
}%
\providecommand \@ifnum [1]{%
 \ifnum #1\expandafter \@firstoftwo
 \else \expandafter \@secondoftwo
 \fi
}%
\providecommand \@ifx [1]{%
 \ifx #1\expandafter \@firstoftwo
 \else \expandafter \@secondoftwo
 \fi
}%
\providecommand \natexlab [1]{#1}%
\providecommand \enquote  [1]{``#1''}%
\providecommand \bibnamefont  [1]{#1}%
\providecommand \bibfnamefont [1]{#1}%
\providecommand \citenamefont [1]{#1}%
\providecommand \href@noop [0]{\@secondoftwo}%
\providecommand \href [0]{\begingroup \@sanitize@url \@href}%
\providecommand \@href[1]{\@@startlink{#1}\@@href}%
\providecommand \@@href[1]{\endgroup#1\@@endlink}%
\providecommand \@sanitize@url [0]{\catcode `\\12\catcode `\$12\catcode
  `\&12\catcode `\#12\catcode `\^12\catcode `\_12\catcode `\%12\relax}%
\providecommand \@@startlink[1]{}%
\providecommand \@@endlink[0]{}%
\providecommand \url  [0]{\begingroup\@sanitize@url \@url }%
\providecommand \@url [1]{\endgroup\@href {#1}{\urlprefix }}%
\providecommand \urlprefix  [0]{URL }%
\providecommand \Eprint [0]{\href }%
\providecommand \doibase [0]{https://doi.org/}%
\providecommand \selectlanguage [0]{\@gobble}%
\providecommand \bibinfo  [0]{\@secondoftwo}%
\providecommand \bibfield  [0]{\@secondoftwo}%
\providecommand \translation [1]{[#1]}%
\providecommand \BibitemOpen [0]{}%
\providecommand \bibitemStop [0]{}%
\providecommand \bibitemNoStop [0]{.\EOS\space}%
\providecommand \EOS [0]{\spacefactor3000\relax}%
\providecommand \BibitemShut  [1]{\csname bibitem#1\endcsname}%
\let\auto@bib@innerbib\@empty
\bibitem [{\citenamefont {Gonçalves}\ \emph {et~al.}(2020)\citenamefont
  {Gonçalves}, \citenamefont {Minassian},\ and\ \citenamefont
  {Melikyan}}]{Gonçalves_2020}%
  \BibitemOpen
  \bibfield  {author} {\bibinfo {author} {\bibfnamefont {M.~R.}\ \bibnamefont
  {Gonçalves}}, \bibinfo {author} {\bibfnamefont {H.}~\bibnamefont
  {Minassian}},\ and\ \bibinfo {author} {\bibfnamefont {A.}~\bibnamefont
  {Melikyan}},\ }\bibfield  {title} {\bibinfo {title} {Plasmonic resonators:
  fundamental properties and applications},\ }\href
  {https://doi.org/10.1088/1361-6463/ab96e9} {\bibfield  {journal} {\bibinfo
  {journal} {Journal of Physics D: Applied Physics}\ }\textbf {\bibinfo
  {volume} {53}},\ \bibinfo {pages} {443002} (\bibinfo {year}
  {2020})}\BibitemShut {NoStop}%
\bibitem [{\citenamefont {Krasnok}\ and\ \citenamefont
  {Alù}(2020)}]{krasnok_active_2020}%
  \BibitemOpen
  \bibfield  {author} {\bibinfo {author} {\bibfnamefont {A.}~\bibnamefont
  {Krasnok}}\ and\ \bibinfo {author} {\bibfnamefont {A.}~\bibnamefont {Alù}},\
  }\bibfield  {title} {\bibinfo {title} {Active nanophotonics},\ }\href
  {https://doi.org/10.1109/JPROC.2020.2985048} {\bibfield  {journal} {\bibinfo
  {journal} {Proceedings of the IEEE}\ }\textbf {\bibinfo {volume} {108}},\
  \bibinfo {pages} {628} (\bibinfo {year} {2020})}\BibitemShut {NoStop}%
\bibitem [{\citenamefont {Pitarke}\ \emph {et~al.}(2006)\citenamefont
  {Pitarke}, \citenamefont {Silkin}, \citenamefont {Chulkov},\ and\
  \citenamefont {Echenique}}]{Pitarke_2007}%
  \BibitemOpen
  \bibfield  {author} {\bibinfo {author} {\bibfnamefont {J.~M.}\ \bibnamefont
  {Pitarke}}, \bibinfo {author} {\bibfnamefont {V.~M.}\ \bibnamefont {Silkin}},
  \bibinfo {author} {\bibfnamefont {E.~V.}\ \bibnamefont {Chulkov}},\ and\
  \bibinfo {author} {\bibfnamefont {P.~M.}\ \bibnamefont {Echenique}},\
  }\bibfield  {title} {\bibinfo {title} {Theory of surface plasmons and
  surface-plasmon polaritons},\ }\href
  {https://doi.org/10.1088/0034-4885/70/1/R01} {\bibfield  {journal} {\bibinfo
  {journal} {Reports on Progress in Physics}\ }\textbf {\bibinfo {volume}
  {70}},\ \bibinfo {pages} {1} (\bibinfo {year} {2006})}\BibitemShut {NoStop}%
\bibitem [{\citenamefont {Wang}\ \emph {et~al.}(2009)\citenamefont {Wang},
  \citenamefont {Wen}, \citenamefont {Yin},\ and\ \citenamefont
  {Wang}}]{Wang:09}%
  \BibitemOpen
  \bibfield  {author} {\bibinfo {author} {\bibfnamefont {T.-B.}\ \bibnamefont
  {Wang}}, \bibinfo {author} {\bibfnamefont {X.-W.}\ \bibnamefont {Wen}},
  \bibinfo {author} {\bibfnamefont {C.-P.}\ \bibnamefont {Yin}},\ and\ \bibinfo
  {author} {\bibfnamefont {H.-Z.}\ \bibnamefont {Wang}},\ }\bibfield  {title}
  {\bibinfo {title} {The transmission characteristics of surface plasmon
  polaritons in ring resonator},\ }\href {https://doi.org/10.1364/OE.17.024096}
  {\bibfield  {journal} {\bibinfo  {journal} {Opt. Express}\ }\textbf {\bibinfo
  {volume} {17}},\ \bibinfo {pages} {24096} (\bibinfo {year}
  {2009})}\BibitemShut {NoStop}%
\bibitem [{\citenamefont {Xiao}\ \emph {et~al.}(2010)\citenamefont {Xiao},
  \citenamefont {Li}, \citenamefont {Jiang}, \citenamefont {Hu}, \citenamefont
  {Li},\ and\ \citenamefont {Gong}}]{Xiao_2010}%
  \BibitemOpen
  \bibfield  {author} {\bibinfo {author} {\bibfnamefont {Y.-F.}\ \bibnamefont
  {Xiao}}, \bibinfo {author} {\bibfnamefont {B.-B.}\ \bibnamefont {Li}},
  \bibinfo {author} {\bibfnamefont {X.}~\bibnamefont {Jiang}}, \bibinfo
  {author} {\bibfnamefont {X.}~\bibnamefont {Hu}}, \bibinfo {author}
  {\bibfnamefont {Y.}~\bibnamefont {Li}},\ and\ \bibinfo {author}
  {\bibfnamefont {Q.}~\bibnamefont {Gong}},\ }\bibfield  {title} {\bibinfo
  {title} {High quality factor, small mode volume, ring-type plasmonic
  microresonator on a silver chip},\ }\href
  {https://doi.org/10.1088/0953-4075/43/3/035402} {\bibfield  {journal}
  {\bibinfo  {journal} {Journal of Physics B: Atomic, Molecular and Optical
  Physics}\ }\textbf {\bibinfo {volume} {43}},\ \bibinfo {pages} {035402}
  (\bibinfo {year} {2010})}\BibitemShut {NoStop}%
\bibitem [{\citenamefont
  {{\O}sterkryger}(2018)}]{8db22ac8bfcc40619fe93ded17fad762}%
  \BibitemOpen
  \bibfield  {author} {\bibinfo {author} {\bibfnamefont {A.}~\bibnamefont
  {{\O}sterkryger}},\ }{\selectlanguage {English}\emph {\bibinfo {title}
  {Engineering of nanophotonic structures for quantum information
  applications.}}},\ \href@noop {} {Ph.D. thesis} (\bibinfo {year}
  {2018})\BibitemShut {NoStop}%
\bibitem [{\citenamefont {Bai}\ \emph {et~al.}(2013)\citenamefont {Bai},
  \citenamefont {Perrin}, \citenamefont {Sauvan}, \citenamefont {Hugonin},\
  and\ \citenamefont {Lalanne}}]{bai_efficient_2013-1}%
  \BibitemOpen
  \bibfield  {author} {\bibinfo {author} {\bibfnamefont {Q.}~\bibnamefont
  {Bai}}, \bibinfo {author} {\bibfnamefont {M.}~\bibnamefont {Perrin}},
  \bibinfo {author} {\bibfnamefont {C.}~\bibnamefont {Sauvan}}, \bibinfo
  {author} {\bibfnamefont {J.-P.}\ \bibnamefont {Hugonin}},\ and\ \bibinfo
  {author} {\bibfnamefont {P.}~\bibnamefont {Lalanne}},\ }\bibfield  {title}
  {{\selectlanguage {EN}\bibinfo {title} {Efficient and intuitive method for
  the analysis of light scattering by a resonant nanostructure}},\ }\href
  {https://doi.org/10.1364/OE.21.027371} {\bibfield  {journal} {\bibinfo
  {journal} {Opt. Express}\ }\textbf {\bibinfo {volume} {21}},\ \bibinfo
  {pages} {27371} (\bibinfo {year} {2013})}\BibitemShut {NoStop}%
\bibitem [{\citenamefont {Hughes}\ \emph {et~al.}(2019)\citenamefont {Hughes},
  \citenamefont {Franke}, \citenamefont {Gustin}, \citenamefont
  {Kamandar~Dezfouli}, \citenamefont {Knorr},\ and\ \citenamefont
  {Richter}}]{Hughes_SPS_2019}%
  \BibitemOpen
  \bibfield  {author} {\bibinfo {author} {\bibfnamefont {S.}~\bibnamefont
  {Hughes}}, \bibinfo {author} {\bibfnamefont {S.}~\bibnamefont {Franke}},
  \bibinfo {author} {\bibfnamefont {C.}~\bibnamefont {Gustin}}, \bibinfo
  {author} {\bibfnamefont {M.}~\bibnamefont {Kamandar~Dezfouli}}, \bibinfo
  {author} {\bibfnamefont {A.}~\bibnamefont {Knorr}},\ and\ \bibinfo {author}
  {\bibfnamefont {M.}~\bibnamefont {Richter}},\ }\bibfield  {title} {\bibinfo
  {title} {Theory and limits of on-demand single-photon sources using plasmonic
  resonators: A quantized quasinormal mode approach},\ }\href
  {https://doi.org/10.1021/acsphotonics.9b00849} {\bibfield  {journal}
  {\bibinfo  {journal} {{ACS} Photonics}\ }\textbf {\bibinfo {volume} {6}},\
  \bibinfo {pages} {2168} (\bibinfo {year} {2019})}\BibitemShut {NoStop}%
\bibitem [{\citenamefont {Chae}\ \emph {et~al.}(2016)\citenamefont {Chae},
  \citenamefont {Lahiri},\ and\ \citenamefont {Centrone}}]{Chae_2016}%
  \BibitemOpen
  \bibfield  {author} {\bibinfo {author} {\bibfnamefont {J.}~\bibnamefont
  {Chae}}, \bibinfo {author} {\bibfnamefont {B.}~\bibnamefont {Lahiri}},\ and\
  \bibinfo {author} {\bibfnamefont {A.}~\bibnamefont {Centrone}},\ }\bibfield
  {title} {\bibinfo {title} {Engineering near-field seira enhancements in
  plasmonic resonators},\ }\href {https://doi.org/10.1021/acsphotonics.5b00466}
  {\bibfield  {journal} {\bibinfo  {journal} {ACS Photonics}\ }\textbf
  {\bibinfo {volume} {3}},\ \bibinfo {pages} {87} (\bibinfo {year}
  {2016})}\BibitemShut {NoStop}%
\bibitem [{\citenamefont {Chikkaraddy}\ \emph {et~al.}(2016)\citenamefont
  {Chikkaraddy}, \citenamefont {de~Nijs}, \citenamefont {Benz}, \citenamefont
  {Barrow}, \citenamefont {Scherman}, \citenamefont {Rosta}, \citenamefont
  {Demetriadou}, \citenamefont {Fox}, \citenamefont {Hess},\ and\ \citenamefont
  {Baumberg}}]{RePEc:nat:nature:v:535:y:2016:i:7610:d:10.1038_nature17974}%
  \BibitemOpen
  \bibfield  {author} {\bibinfo {author} {\bibfnamefont {R.}~\bibnamefont
  {Chikkaraddy}}, \bibinfo {author} {\bibfnamefont {B.}~\bibnamefont
  {de~Nijs}}, \bibinfo {author} {\bibfnamefont {F.}~\bibnamefont {Benz}},
  \bibinfo {author} {\bibfnamefont {S.~J.}\ \bibnamefont {Barrow}}, \bibinfo
  {author} {\bibfnamefont {O.~A.}\ \bibnamefont {Scherman}}, \bibinfo {author}
  {\bibfnamefont {E.}~\bibnamefont {Rosta}}, \bibinfo {author} {\bibfnamefont
  {A.}~\bibnamefont {Demetriadou}}, \bibinfo {author} {\bibfnamefont
  {P.}~\bibnamefont {Fox}}, \bibinfo {author} {\bibfnamefont {O.}~\bibnamefont
  {Hess}},\ and\ \bibinfo {author} {\bibfnamefont {J.~J.}\ \bibnamefont
  {Baumberg}},\ }\bibfield  {title} {\bibinfo {title} {{Single-molecule strong
  coupling at room temperature in plasmonic nanocavities}},\ }\href
  {https://doi.org/10.1038/nature17974} {\bibfield  {journal} {\bibinfo
  {journal} {Nature}\ }\textbf {\bibinfo {volume} {535}},\ \bibinfo {pages}
  {127} (\bibinfo {year} {2016})}\BibitemShut {NoStop}%
\bibitem [{\citenamefont {Koenderink}(2010)}]{Koenderink2010}%
  \BibitemOpen
  \bibfield  {author} {\bibinfo {author} {\bibfnamefont {A.~F.}\ \bibnamefont
  {Koenderink}},\ }\bibfield  {title} {\bibinfo {title} {On the use of purcell
  factors for plasmon antennas},\ }\href {https://doi.org/10.1364/ol.35.004208}
  {\bibfield  {journal} {\bibinfo  {journal} {Optics Letters}\ }\textbf
  {\bibinfo {volume} {35}},\ \bibinfo {pages} {4208} (\bibinfo {year}
  {2010})}\BibitemShut {NoStop}%
\bibitem [{\citenamefont {Kristensen}\ and\ \citenamefont
  {Hughes}(2014)}]{kristensen_modes_2014}%
  \BibitemOpen
  \bibfield  {author} {\bibinfo {author} {\bibfnamefont {P.~T.}\ \bibnamefont
  {Kristensen}}\ and\ \bibinfo {author} {\bibfnamefont {S.}~\bibnamefont
  {Hughes}},\ }\bibfield  {title} {{\selectlanguage {en}\bibinfo {title} {Modes
  and {Mode} {Volumes} of {Leaky} {Optical} {Cavities} and {Plasmonic}
  {Nanoresonators}}},\ }\href {https://doi.org/10.1021/ph400114e} {\bibfield
  {journal} {\bibinfo  {journal} {ACS Photonics}\ }\textbf {\bibinfo {volume}
  {1}},\ \bibinfo {pages} {2} (\bibinfo {year} {2014})}\BibitemShut {NoStop}%
\bibitem [{\citenamefont {Kristensen}\ \emph {et~al.}(2020)\citenamefont
  {Kristensen}, \citenamefont {Herrmann}, \citenamefont {Intravaia},\ and\
  \citenamefont {Busch}}]{kristensen_modeling_2020}%
  \BibitemOpen
  \bibfield  {author} {\bibinfo {author} {\bibfnamefont {P.~T.}\ \bibnamefont
  {Kristensen}}, \bibinfo {author} {\bibfnamefont {K.}~\bibnamefont
  {Herrmann}}, \bibinfo {author} {\bibfnamefont {F.}~\bibnamefont
  {Intravaia}},\ and\ \bibinfo {author} {\bibfnamefont {K.}~\bibnamefont
  {Busch}},\ }\bibfield  {title} {\bibinfo {title} {Modeling electromagnetic
  resonators using quasinormal modes},\ }\href
  {https://doi.org/10.1364/AOP.377940} {\bibfield  {journal} {\bibinfo
  {journal} {Adv. Opt. Photonics}\ }\textbf {\bibinfo {volume} {12}},\ \bibinfo
  {pages} {612} (\bibinfo {year} {2020})}\BibitemShut {NoStop}%
\bibitem [{\citenamefont {Ren}\ \emph {et~al.}(2021)\citenamefont {Ren},
  \citenamefont {Franke},\ and\ \citenamefont {Hughes}}]{ren_quasinormal_2021}%
  \BibitemOpen
  \bibfield  {author} {\bibinfo {author} {\bibfnamefont {J.}~\bibnamefont
  {Ren}}, \bibinfo {author} {\bibfnamefont {S.}~\bibnamefont {Franke}},\ and\
  \bibinfo {author} {\bibfnamefont {S.}~\bibnamefont {Hughes}},\ }\bibfield
  {title} {\bibinfo {title} {Quasinormal modes, local density of states, and
  {Classical Purcell} factors for coupled loss-gain resonators},\ }\href
  {https://doi.org/10.1103/PhysRevX.11.041020} {\bibfield  {journal} {\bibinfo
  {journal} {Phys. Rev. X}\ }\textbf {\bibinfo {volume} {11}},\ \bibinfo
  {pages} {041020} (\bibinfo {year} {2021})}\BibitemShut {NoStop}%
\bibitem [{\citenamefont {Lalanne}\ \emph {et~al.}(2018)\citenamefont
  {Lalanne}, \citenamefont {Yan}, \citenamefont {Vynck}, \citenamefont
  {Sauvan},\ and\ \citenamefont {Hugonin}}]{lalanne_light_2018}%
  \BibitemOpen
  \bibfield  {author} {\bibinfo {author} {\bibfnamefont {P.}~\bibnamefont
  {Lalanne}}, \bibinfo {author} {\bibfnamefont {W.}~\bibnamefont {Yan}},
  \bibinfo {author} {\bibfnamefont {K.}~\bibnamefont {Vynck}}, \bibinfo
  {author} {\bibfnamefont {C.}~\bibnamefont {Sauvan}},\ and\ \bibinfo {author}
  {\bibfnamefont {J.-P.}\ \bibnamefont {Hugonin}},\ }\bibfield  {title}
  {\bibinfo {title} {Light interaction with photonic and plasmonic
  resonances},\ }\href {https://doi.org/10.1002/lpor.201700113} {\bibfield
  {journal} {\bibinfo  {journal} {Laser Photonics Rev.}\ }\textbf {\bibinfo
  {volume} {12}},\ \bibinfo {pages} {1700113} (\bibinfo {year}
  {2018})}\BibitemShut {NoStop}%
\bibitem [{\citenamefont {Perrin}\ \emph {et~al.}(2016)\citenamefont {Perrin},
  \citenamefont {Yang},\ and\ \citenamefont {Lalanne}}]{perrin_QNM_2016}%
  \BibitemOpen
  \bibfield  {author} {\bibinfo {author} {\bibfnamefont {M.}~\bibnamefont
  {Perrin}}, \bibinfo {author} {\bibfnamefont {J.}~\bibnamefont {Yang}},\ and\
  \bibinfo {author} {\bibfnamefont {P.}~\bibnamefont {Lalanne}},\ }\bibfield
  {title} {\bibinfo {title} {{Analytical treatment of the interaction between
  light, plasmonic and quantum resonances: quasi-normal mode expansion}},\ }in\
  \href {https://doi.org/10.1117/12.2209707} {\emph {\bibinfo {booktitle}
  {Quantum Sensing and Nano Electronics and Photonics XIII}}},\ Vol.\ \bibinfo
  {volume} {9755},\ \bibinfo {editor} {edited by\ \bibinfo {editor}
  {\bibfnamefont {M.}~\bibnamefont {Razeghi}}},\ \bibinfo {organization}
  {International Society for Optics and Photonics}\ (\bibinfo  {publisher}
  {SPIE},\ \bibinfo {year} {2016})\ p.\ \bibinfo {pages} {97551J}\BibitemShut
  {NoStop}%
\bibitem [{\citenamefont {Franke}\ \emph {et~al.}(2019)\citenamefont {Franke},
  \citenamefont {Hughes}, \citenamefont {Dezfouli}, \citenamefont {Kristensen},
  \citenamefont {Busch}, \citenamefont {Knorr},\ and\ \citenamefont
  {Richter}}]{franke_quantization_2019}%
  \BibitemOpen
  \bibfield  {author} {\bibinfo {author} {\bibfnamefont {S.}~\bibnamefont
  {Franke}}, \bibinfo {author} {\bibfnamefont {S.}~\bibnamefont {Hughes}},
  \bibinfo {author} {\bibfnamefont {M.~K.}\ \bibnamefont {Dezfouli}}, \bibinfo
  {author} {\bibfnamefont {P.~T.}\ \bibnamefont {Kristensen}}, \bibinfo
  {author} {\bibfnamefont {K.}~\bibnamefont {Busch}}, \bibinfo {author}
  {\bibfnamefont {A.}~\bibnamefont {Knorr}},\ and\ \bibinfo {author}
  {\bibfnamefont {M.}~\bibnamefont {Richter}},\ }\bibfield  {title} {\bibinfo
  {title} {Quantization of quasinormal modes for open cavities and plasmonic
  cavity quantum electrodynamics},\ }\href
  {https://doi.org/10.1103/PhysRevLett.122.213901} {\bibfield  {journal}
  {\bibinfo  {journal} {Phys. Rev. Lett.}\ }\textbf {\bibinfo {volume} {122}},\
  \bibinfo {pages} {213901} (\bibinfo {year} {2019})}\BibitemShut {NoStop}%
\bibitem [{\citenamefont {Van~Vlack}()}]{vlack_dyadic_2012}%
  \BibitemOpen
  \bibfield  {author} {\bibinfo {author} {\bibfnamefont {C.~P.}\ \bibnamefont
  {Van~Vlack}},\ }\href {https://qspace.library.queensu.ca/handle/1974/7128}
  {\bibinfo {title} {Dyadic {Green} functions and their applications in
  classical and quantum nanophotonics}},\ \bibinfo {note} {phD Thesis, Queen's
  University (2012)}\BibitemShut {NoStop}%
\bibitem [{\citenamefont {Ge}\ \emph {et~al.}(2014)\citenamefont {Ge},
  \citenamefont {Kristensen}, \citenamefont {Young},\ and\ \citenamefont
  {Hughes}}]{ge_quasinormal_2014}%
  \BibitemOpen
  \bibfield  {author} {\bibinfo {author} {\bibfnamefont {R.-C.}\ \bibnamefont
  {Ge}}, \bibinfo {author} {\bibfnamefont {P.~T.}\ \bibnamefont {Kristensen}},
  \bibinfo {author} {\bibfnamefont {J.~F.}\ \bibnamefont {Young}},\ and\
  \bibinfo {author} {\bibfnamefont {S.}~\bibnamefont {Hughes}},\ }\bibfield
  {title} {\bibinfo {title} {Quasinormal mode approach to modelling
  light-emission and propagation in nanoplasmonics},\ }\href
  {https://doi.org/10.1088/1367-2630/16/11/113048} {\bibfield  {journal}
  {\bibinfo  {journal} {New J. Phys.}\ }\textbf {\bibinfo {volume} {16}},\
  \bibinfo {pages} {113048} (\bibinfo {year} {2014})}\BibitemShut {NoStop}%
\bibitem [{\citenamefont {Dung}\ \emph {et~al.}(1998)\citenamefont {Dung},
  \citenamefont {Kn\"{o}ll},\ and\ \citenamefont
  {Welsch}}]{dung_three-dimensional_1998}%
  \BibitemOpen
  \bibfield  {author} {\bibinfo {author} {\bibfnamefont {H.~T.}\ \bibnamefont
  {Dung}}, \bibinfo {author} {\bibfnamefont {L.}~\bibnamefont {Kn\"{o}ll}},\
  and\ \bibinfo {author} {\bibfnamefont {D.-G.}\ \bibnamefont {Welsch}},\
  }\bibfield  {title} {\bibinfo {title} {Three-dimensional quantization of the
  electromagnetic field in dispersive and absorbing inhomogeneous
  dielectrics},\ }\href {https://doi.org/10.1103/PhysRevA.57.3931} {\bibfield
  {journal} {\bibinfo  {journal} {Phys. Rev. A}\ }\textbf {\bibinfo {volume}
  {57}},\ \bibinfo {pages} {3931} (\bibinfo {year} {1998})}\BibitemShut
  {NoStop}%
\bibitem [{\citenamefont {Righini}\ \emph {et~al.}(2007)\citenamefont
  {Righini}, \citenamefont {Zelenina},\ and\ \citenamefont
  {Girard}}]{Righini_2007_trapping}%
  \BibitemOpen
  \bibfield  {author} {\bibinfo {author} {\bibfnamefont {M.}~\bibnamefont
  {Righini}}, \bibinfo {author} {\bibfnamefont {A.}~\bibnamefont {Zelenina}},\
  and\ \bibinfo {author} {\bibfnamefont {C.~e.~a.}\ \bibnamefont {Girard}},\
  }\bibfield  {title} {\bibinfo {title} {Parallel and selective trapping in a
  patterned plasmonic landscape},\ }\href {https://doi.org/10.1038/nphys624}
  {\bibfield  {journal} {\bibinfo  {journal} {Nature Physics}\ }\textbf
  {\bibinfo {volume} {3}},\ \bibinfo {pages} {477–480} (\bibinfo {year}
  {2007})}\BibitemShut {NoStop}%
\bibitem [{\citenamefont {Sharma}\ and\ \citenamefont
  {Shrivastava}(2008)}]{sharma_2018_dielectric}%
  \BibitemOpen
  \bibfield  {author} {\bibinfo {author} {\bibfnamefont {A.}~\bibnamefont
  {Sharma}}\ and\ \bibinfo {author} {\bibfnamefont {S.~C.}\ \bibnamefont
  {Shrivastava}},\ }\bibfield  {title} {\bibinfo {title} {Analysis of resonant
  frequency and quality factor of dielectric resonator at different dielectric
  constant materials},\ }in\ \href {https://doi.org/10.1109/AMTA.2008.4763084}
  {\emph {\bibinfo {booktitle} {2008 International Conference on Recent
  Advances in Microwave Theory and Applications}}}\ (\bibinfo {year} {2008})\
  pp.\ \bibinfo {pages} {593--595}\BibitemShut {NoStop}%
\bibitem [{\citenamefont {Cogn\'{e}e}\ \emph {et~al.}(2019)\citenamefont
  {Cogn\'{e}e}, \citenamefont {Doeleman}, \citenamefont {Lalanne},\ and\
  \citenamefont {Koenderink}}]{cognee_cooperative_2019}%
  \BibitemOpen
  \bibfield  {author} {\bibinfo {author} {\bibfnamefont {K.~G.}\ \bibnamefont
  {Cogn\'{e}e}}, \bibinfo {author} {\bibfnamefont {H.~M.}\ \bibnamefont
  {Doeleman}}, \bibinfo {author} {\bibfnamefont {P.}~\bibnamefont {Lalanne}},\
  and\ \bibinfo {author} {\bibfnamefont {A.~F.}\ \bibnamefont {Koenderink}},\
  }\bibfield  {title} {\bibinfo {title} {Cooperative interactions between
  nano-antennas in a high-{Q} cavity for unidirectional light sources},\ }\href
  {https://doi.org/10.1038/s41377-019-0227-x} {\bibfield  {journal} {\bibinfo
  {journal} {Light: Sci. Appl.}\ }\textbf {\bibinfo {volume} {8}},\ \bibinfo
  {pages} {115} (\bibinfo {year} {2019})}\BibitemShut {NoStop}%
\bibitem [{\citenamefont {Melli}\ \emph {et~al.}(2013)\citenamefont {Melli},
  \citenamefont {Polyakov}, \citenamefont {Gargas}, \citenamefont {Huynh},
  \citenamefont {Scipioni}, \citenamefont {Bao}, \citenamefont {Ogletree},
  \citenamefont {Schuck}, \citenamefont {Cabrini},\ and\ \citenamefont
  {Weber-Bargioni}}]{melli_2013_Qlim}%
  \BibitemOpen
  \bibfield  {author} {\bibinfo {author} {\bibfnamefont {M.}~\bibnamefont
  {Melli}}, \bibinfo {author} {\bibfnamefont {A.}~\bibnamefont {Polyakov}},
  \bibinfo {author} {\bibfnamefont {D.}~\bibnamefont {Gargas}}, \bibinfo
  {author} {\bibfnamefont {C.}~\bibnamefont {Huynh}}, \bibinfo {author}
  {\bibfnamefont {L.}~\bibnamefont {Scipioni}}, \bibinfo {author}
  {\bibfnamefont {W.}~\bibnamefont {Bao}}, \bibinfo {author} {\bibfnamefont
  {D.~F.}\ \bibnamefont {Ogletree}}, \bibinfo {author} {\bibfnamefont {P.~J.}\
  \bibnamefont {Schuck}}, \bibinfo {author} {\bibfnamefont {S.}~\bibnamefont
  {Cabrini}},\ and\ \bibinfo {author} {\bibfnamefont {A.}~\bibnamefont
  {Weber-Bargioni}},\ }\bibfield  {title} {\bibinfo {title} {Reaching the
  theoretical resonance quality factor limit in coaxial plasmonic
  nanoresonators fabricated by helium ion lithography},\ }\href
  {https://doi.org/10.1021/nl400844a} {\bibfield  {journal} {\bibinfo
  {journal} {Nano Letters}\ }\textbf {\bibinfo {volume} {13}},\ \bibinfo
  {pages} {2687} (\bibinfo {year} {2013})}\BibitemShut {NoStop}%
\bibitem [{\citenamefont {Lilley}\ \emph {et~al.}(2015)\citenamefont {Lilley},
  \citenamefont {Messner},\ and\ \citenamefont {Unterrainer}}]{Lilley:15}%
  \BibitemOpen
  \bibfield  {author} {\bibinfo {author} {\bibfnamefont {G.}~\bibnamefont
  {Lilley}}, \bibinfo {author} {\bibfnamefont {M.}~\bibnamefont {Messner}},\
  and\ \bibinfo {author} {\bibfnamefont {K.}~\bibnamefont {Unterrainer}},\
  }\bibfield  {title} {\bibinfo {title} {Improving the quality factor of the
  localized surface plasmon resonance},\ }\href
  {https://doi.org/10.1364/OME.5.002112} {\bibfield  {journal} {\bibinfo
  {journal} {Opt. Mater. Express}\ }\textbf {\bibinfo {volume} {5}},\ \bibinfo
  {pages} {2112} (\bibinfo {year} {2015})}\BibitemShut {NoStop}%
\bibitem [{\citenamefont {Kuttge}\ \emph {et~al.}(2010)\citenamefont {Kuttge},
  \citenamefont {García~de Abajo},\ and\ \citenamefont
  {Polman}}]{kuttge_nanodisk_2010}%
  \BibitemOpen
  \bibfield  {author} {\bibinfo {author} {\bibfnamefont {M.}~\bibnamefont
  {Kuttge}}, \bibinfo {author} {\bibfnamefont {F.~J.}\ \bibnamefont {García~de
  Abajo}},\ and\ \bibinfo {author} {\bibfnamefont {A.}~\bibnamefont {Polman}},\
  }\bibfield  {title} {\bibinfo {title} {Ultrasmall mode volume plasmonic
  nanodisk resonators},\ }\href {https://doi.org/10.1021/nl902546r} {\bibfield
  {journal} {\bibinfo  {journal} {Nano Letters}\ }\textbf {\bibinfo {volume}
  {10}},\ \bibinfo {pages} {1537} (\bibinfo {year} {2010})},\ \bibinfo {note}
  {pMID: 19813755}\BibitemShut {NoStop}%
\bibitem [{\citenamefont {Ren}\ \emph {et~al.}(2020)\citenamefont {Ren},
  \citenamefont {Franke}, \citenamefont {Knorr}, \citenamefont {Richter},\ and\
  \citenamefont {Hughes}}]{ren_near-field_2020}%
  \BibitemOpen
  \bibfield  {author} {\bibinfo {author} {\bibfnamefont {J.}~\bibnamefont
  {Ren}}, \bibinfo {author} {\bibfnamefont {S.}~\bibnamefont {Franke}},
  \bibinfo {author} {\bibfnamefont {A.}~\bibnamefont {Knorr}}, \bibinfo
  {author} {\bibfnamefont {M.}~\bibnamefont {Richter}},\ and\ \bibinfo {author}
  {\bibfnamefont {S.}~\bibnamefont {Hughes}},\ }\bibfield  {title} {\bibinfo
  {title} {Near-field to far-field transformations of optical quasinormal modes
  and efficient calculation of quantized quasinormal modes for open cavities
  and plasmonic resonators},\ }\href
  {https://doi.org/10.1103/PhysRevB.101.205402} {\bibfield  {journal} {\bibinfo
   {journal} {Phys. Rev. B}\ }\textbf {\bibinfo {volume} {101}},\ \bibinfo
  {pages} {205402} (\bibinfo {year} {2020})}\BibitemShut {NoStop}%
\bibitem [{\citenamefont {Stockman}(2008)}]{stockman_spasers_2008}%
  \BibitemOpen
  \bibfield  {author} {\bibinfo {author} {\bibfnamefont {M.~I.}\ \bibnamefont
  {Stockman}},\ }\bibfield  {title} {\bibinfo {title} {Spasers explained},\
  }\href {https://doi.org/10.1038/nphoton.2008.85} {\bibfield  {journal}
  {\bibinfo  {journal} {Nature Photonics}\ }\textbf {\bibinfo {volume} {2}},\
  \bibinfo {pages} {327} (\bibinfo {year} {2008})}\BibitemShut {NoStop}%
\bibitem [{\citenamefont
  {Stockman}(2011{\natexlab{a}})}]{stockman_spaser_2011}%
  \BibitemOpen
  \bibfield  {author} {\bibinfo {author} {\bibfnamefont {M.~I.}\ \bibnamefont
  {Stockman}},\ }\bibfield  {title} {\bibinfo {title} {Spaser action, loss
  compensation, and stability in plasmonic systems with gain},\ }\href
  {https://doi.org/10.1103/PhysRevLett.106.156802} {\bibfield  {journal}
  {\bibinfo  {journal} {Physical Review Letters}\ }\textbf {\bibinfo {volume}
  {106}},\ \bibinfo {pages} {156802} (\bibinfo {year}
  {2011}{\natexlab{a}})}\BibitemShut {NoStop}%
\bibitem [{\citenamefont {Stockman}(2011{\natexlab{b}})}]{stockman_gain_2011}%
  \BibitemOpen
  \bibfield  {author} {\bibinfo {author} {\bibfnamefont {M.~I.}\ \bibnamefont
  {Stockman}},\ }\bibfield  {title} {\bibinfo {title} {Loss compensation by
  gain and spasing},\ }\href {https://doi.org/10.1098/rsta.2011.0143}
  {\bibfield  {journal} {\bibinfo  {journal} {Philosophical Transactions of the
  Royal Society A: Mathematical, Physical and Engineering Sciences}\ }\textbf
  {\bibinfo {volume} {369}},\ \bibinfo {pages} {3510} (\bibinfo {year}
  {2011}{\natexlab{b}})}\BibitemShut {NoStop}%
\bibitem [{\citenamefont {Russev}\ \emph {et~al.}(2012)\citenamefont {Russev},
  \citenamefont {Tsutsumanova},\ and\ \citenamefont
  {Tzonev}}]{russev_conditions_2012}%
  \BibitemOpen
  \bibfield  {author} {\bibinfo {author} {\bibfnamefont {S.~C.}\ \bibnamefont
  {Russev}}, \bibinfo {author} {\bibfnamefont {G.~G.}\ \bibnamefont
  {Tsutsumanova}},\ and\ \bibinfo {author} {\bibfnamefont {A.~N.}\ \bibnamefont
  {Tzonev}},\ }\bibfield  {title} {\bibinfo {title} {Conditions for loss
  compensation of surface plasmon polaritons propagation on a metal/gain medium
  boundary},\ }\href {https://doi.org/10.1007/s11468-011-9288-2} {\bibfield
  {journal} {\bibinfo  {journal} {Plasmonics}\ }\textbf {\bibinfo {volume}
  {7}},\ \bibinfo {pages} {151} (\bibinfo {year} {2012})}\BibitemShut {NoStop}%
\bibitem [{\citenamefont {Veltri}\ and\ \citenamefont
  {Aradian}(2012)}]{veltri_PRB_2012}%
  \BibitemOpen
  \bibfield  {author} {\bibinfo {author} {\bibfnamefont {A.}~\bibnamefont
  {Veltri}}\ and\ \bibinfo {author} {\bibfnamefont {A.}~\bibnamefont
  {Aradian}},\ }\bibfield  {title} {\bibinfo {title} {Optical response of a
  metallic nanoparticle immersed in a medium with optical gain},\ }\href
  {https://doi.org/10.1103/PhysRevB.85.115429} {\bibfield  {journal} {\bibinfo
  {journal} {Phys. Rev. B}\ }\textbf {\bibinfo {volume} {85}},\ \bibinfo
  {pages} {115429} (\bibinfo {year} {2012})}\BibitemShut {NoStop}%
\bibitem [{\citenamefont {De~Leon}\ and\ \citenamefont
  {Berini}(2010)}]{De_Leon_Berini_2010}%
  \BibitemOpen
  \bibfield  {author} {\bibinfo {author} {\bibfnamefont {I.}~\bibnamefont
  {De~Leon}}\ and\ \bibinfo {author} {\bibfnamefont {P.}~\bibnamefont
  {Berini}},\ }\bibfield  {title} {\bibinfo {title} {Amplification of
  long-range surface plasmons by a dipolar gain medium},\ }\href
  {https://doi.org/10.1038/nphoton.2010.37} {\bibfield  {journal} {\bibinfo
  {journal} {Nature Photonics}\ }\textbf {\bibinfo {volume} {4}},\ \bibinfo
  {pages} {382–387} (\bibinfo {year} {2010})}\BibitemShut {NoStop}%
\bibitem [{\citenamefont {Noginov}(2008)}]{Noginov_2008}%
  \BibitemOpen
  \bibfield  {author} {\bibinfo {author} {\bibfnamefont {M.~A.}\ \bibnamefont
  {Noginov}},\ }\bibfield  {title} {\bibinfo {title} {{Compensation of surface
  plasmon loss by gain in dielectric medium}},\ }\href
  {https://doi.org/10.1117/1.3073670} {\bibfield  {journal} {\bibinfo
  {journal} {Journal of Nanophotonics}\ }\textbf {\bibinfo {volume} {2}},\
  \bibinfo {pages} {021855} (\bibinfo {year} {2008})}\BibitemShut {NoStop}%
\bibitem [{\citenamefont {Berini}\ and\ \citenamefont
  {De~Leon}(2011)}]{Berini_De_Leon_2011}%
  \BibitemOpen
  \bibfield  {author} {\bibinfo {author} {\bibfnamefont {P.}~\bibnamefont
  {Berini}}\ and\ \bibinfo {author} {\bibfnamefont {I.}~\bibnamefont
  {De~Leon}},\ }\bibfield  {title} {\bibinfo {title} {Surface
  plasmon–polariton amplifiers and lasers},\ }\href
  {https://doi.org/10.1038/nphoton.2011.285} {\bibfield  {journal} {\bibinfo
  {journal} {Nature Photonics}\ }\textbf {\bibinfo {volume} {6}},\ \bibinfo
  {pages} {16–24} (\bibinfo {year} {2011})}\BibitemShut {NoStop}%
\bibitem [{\citenamefont {Pustovit}\ \emph {et~al.}(2015)\citenamefont
  {Pustovit}, \citenamefont {Capolino},\ and\ \citenamefont
  {Aradian}}]{Pustovit:15}%
  \BibitemOpen
  \bibfield  {author} {\bibinfo {author} {\bibfnamefont {V.}~\bibnamefont
  {Pustovit}}, \bibinfo {author} {\bibfnamefont {F.}~\bibnamefont {Capolino}},\
  and\ \bibinfo {author} {\bibfnamefont {A.}~\bibnamefont {Aradian}},\
  }\bibfield  {title} {\bibinfo {title} {Cooperative plasmon-mediated effects
  and loss compensation by gain dyes near a metal nanoparticle},\ }\href
  {https://doi.org/10.1364/JOSAB.32.000188} {\bibfield  {journal} {\bibinfo
  {journal} {J. Opt. Soc. Am. B}\ }\textbf {\bibinfo {volume} {32}},\ \bibinfo
  {pages} {188} (\bibinfo {year} {2015})}\BibitemShut {NoStop}%
\bibitem [{\citenamefont {Liu}\ \emph {et~al.}(2011)\citenamefont {Liu},
  \citenamefont {Li}, \citenamefont {Zhou}, \citenamefont {Gan},\ and\
  \citenamefont {Li}}]{liu_efficient_2011}%
  \BibitemOpen
  \bibfield  {author} {\bibinfo {author} {\bibfnamefont {S.-Y.}\ \bibnamefont
  {Liu}}, \bibinfo {author} {\bibfnamefont {J.}~\bibnamefont {Li}}, \bibinfo
  {author} {\bibfnamefont {F.}~\bibnamefont {Zhou}}, \bibinfo {author}
  {\bibfnamefont {L.}~\bibnamefont {Gan}},\ and\ \bibinfo {author}
  {\bibfnamefont {Z.-Y.}\ \bibnamefont {Li}},\ }\bibfield  {title} {\bibinfo
  {title} {Efficient surface plasmon amplification from gain-assisted gold
  nanorods},\ }\href {https://doi.org/10.1364/OL.36.001296} {\bibfield
  {journal} {\bibinfo  {journal} {Optics Letters}\ }\textbf {\bibinfo {volume}
  {36}},\ \bibinfo {pages} {1296} (\bibinfo {year} {2011})}\BibitemShut
  {NoStop}%
\bibitem [{\citenamefont {Cai}\ \emph {et~al.}(2018)\citenamefont {Cai},
  \citenamefont {Zhou}, \citenamefont {Zhang},\ and\ \citenamefont
  {Li}}]{Cai:18}%
  \BibitemOpen
  \bibfield  {author} {\bibinfo {author} {\bibfnamefont {J.}~\bibnamefont
  {Cai}}, \bibinfo {author} {\bibfnamefont {Y.~J.}\ \bibnamefont {Zhou}},
  \bibinfo {author} {\bibfnamefont {Y.}~\bibnamefont {Zhang}},\ and\ \bibinfo
  {author} {\bibfnamefont {Q.~Y.}\ \bibnamefont {Li}},\ }\bibfield  {title}
  {\bibinfo {title} {Gain-assisted ultra-high-q spoof plasmonic resonator for
  the sensing of polar liquids},\ }\href {https://doi.org/10.1364/OE.26.025460}
  {\bibfield  {journal} {\bibinfo  {journal} {Opt. Express}\ }\textbf {\bibinfo
  {volume} {26}},\ \bibinfo {pages} {25460} (\bibinfo {year}
  {2018})}\BibitemShut {NoStop}%
\bibitem [{\citenamefont {Fang}\ \emph {et~al.}(2011)\citenamefont {Fang},
  \citenamefont {Huang}, \citenamefont {Koschny},\ and\ \citenamefont
  {Soukoulis}}]{Fang:11}%
  \BibitemOpen
  \bibfield  {author} {\bibinfo {author} {\bibfnamefont {A.}~\bibnamefont
  {Fang}}, \bibinfo {author} {\bibfnamefont {Z.}~\bibnamefont {Huang}},
  \bibinfo {author} {\bibfnamefont {T.}~\bibnamefont {Koschny}},\ and\ \bibinfo
  {author} {\bibfnamefont {C.~M.}\ \bibnamefont {Soukoulis}},\ }\bibfield
  {title} {\bibinfo {title} {Overcoming the losses of a split ring resonator
  array with gain},\ }\href {https://doi.org/10.1364/OE.19.012688} {\bibfield
  {journal} {\bibinfo  {journal} {Opt. Express}\ }\textbf {\bibinfo {volume}
  {19}},\ \bibinfo {pages} {12688} (\bibinfo {year} {2011})}\BibitemShut
  {NoStop}%
\bibitem [{\citenamefont {Ding}\ \emph {et~al.}(2013)\citenamefont {Ding},
  \citenamefont {He}, \citenamefont {Wang}, \citenamefont {Fan}, \citenamefont
  {Cai},\ and\ \citenamefont {Liang}}]{Ding_2013}%
  \BibitemOpen
  \bibfield  {author} {\bibinfo {author} {\bibfnamefont {P.}~\bibnamefont
  {Ding}}, \bibinfo {author} {\bibfnamefont {J.}~\bibnamefont {He}}, \bibinfo
  {author} {\bibfnamefont {J.}~\bibnamefont {Wang}}, \bibinfo {author}
  {\bibfnamefont {C.}~\bibnamefont {Fan}}, \bibinfo {author} {\bibfnamefont
  {G.}~\bibnamefont {Cai}},\ and\ \bibinfo {author} {\bibfnamefont
  {E.}~\bibnamefont {Liang}},\ }\bibfield  {title} {\bibinfo {title}
  {Low-threshold surface plasmon amplification from a gain-assisted
  core–shell nanoparticle with broken symmetry},\ }\href
  {https://doi.org/10.1088/2040-8978/15/10/105001} {\bibfield  {journal}
  {\bibinfo  {journal} {Journal of Optics}\ }\textbf {\bibinfo {volume} {15}},\
  \bibinfo {pages} {105001} (\bibinfo {year} {2013})}\BibitemShut {NoStop}%
\bibitem [{\citenamefont {Wang}\ \emph {et~al.}(2017)\citenamefont {Wang},
  \citenamefont {Meng}, \citenamefont {Kildishev}, \citenamefont {Boltasseva},\
  and\ \citenamefont {Shalaev}}]{Wang_nanolaser_2017}%
  \BibitemOpen
  \bibfield  {author} {\bibinfo {author} {\bibfnamefont {Z.}~\bibnamefont
  {Wang}}, \bibinfo {author} {\bibfnamefont {X.}~\bibnamefont {Meng}}, \bibinfo
  {author} {\bibfnamefont {A.~V.}\ \bibnamefont {Kildishev}}, \bibinfo {author}
  {\bibfnamefont {A.}~\bibnamefont {Boltasseva}},\ and\ \bibinfo {author}
  {\bibfnamefont {V.~M.}\ \bibnamefont {Shalaev}},\ }\bibfield  {title}
  {\bibinfo {title} {Nanolasers enabled by metallic nanoparticles: From spasers
  to random lasers},\ }\href {https://doi.org/10.1002/lpor.201700212}
  {\bibfield  {journal} {\bibinfo  {journal} {Laser {\&} Photonics Reviews}\
  }\textbf {\bibinfo {volume} {11}},\ \bibinfo {pages} {1700212} (\bibinfo
  {year} {2017})}\BibitemShut {NoStop}%
\bibitem [{\citenamefont {Zhong}\ and\ \citenamefont
  {Li}(2013)}]{Zhong_spaser_2013}%
  \BibitemOpen
  \bibfield  {author} {\bibinfo {author} {\bibfnamefont {X.-L.}\ \bibnamefont
  {Zhong}}\ and\ \bibinfo {author} {\bibfnamefont {Z.-Y.}\ \bibnamefont {Li}},\
  }\bibfield  {title} {\bibinfo {title} {All-analytical semiclassical theory of
  spaser performance in a plasmonic nanocavity},\ }\href
  {https://doi.org/10.1103/PhysRevB.88.085101} {\bibfield  {journal} {\bibinfo
  {journal} {Phys. Rev. B}\ }\textbf {\bibinfo {volume} {88}},\ \bibinfo
  {pages} {085101} (\bibinfo {year} {2013})}\BibitemShut {NoStop}%
\bibitem [{\citenamefont {Zhong}\ \emph {et~al.}(2013)\citenamefont {Zhong},
  \citenamefont {Hong},\ and\ \citenamefont {Li}}]{Zhong_Hong_Li_2013}%
  \BibitemOpen
  \bibfield  {author} {\bibinfo {author} {\bibfnamefont {X.~L.}\ \bibnamefont
  {Zhong}}, \bibinfo {author} {\bibfnamefont {M.~H.}\ \bibnamefont {Hong}},\
  and\ \bibinfo {author} {\bibfnamefont {Z.~Y.}\ \bibnamefont {Li}},\
  }\bibfield  {title} {\bibinfo {title} {Spaser in plasmonic nano-antenna
  evaluated by an analytical theory},\ }\href
  {https://doi.org/10.1007/s00339-013-7926-6} {\bibfield  {journal} {\bibinfo
  {journal} {Applied Physics A}\ }\textbf {\bibinfo {volume} {115}},\ \bibinfo
  {pages} {5–11} (\bibinfo {year} {2013})}\BibitemShut {NoStop}%
\bibitem [{\citenamefont {Warnakula}\ \emph {et~al.}(2018)\citenamefont
  {Warnakula}, \citenamefont {Stockman},\ and\ \citenamefont
  {Premaratne}}]{Warnakula_2018}%
  \BibitemOpen
  \bibfield  {author} {\bibinfo {author} {\bibfnamefont {T.}~\bibnamefont
  {Warnakula}}, \bibinfo {author} {\bibfnamefont {M.~I.}\ \bibnamefont
  {Stockman}},\ and\ \bibinfo {author} {\bibfnamefont {M.}~\bibnamefont
  {Premaratne}},\ }\bibfield  {title} {\bibinfo {title} {Improved scheme for
  modeling a spaser made of identical gain elements},\ }\href
  {https://doi.org/10.1364/JOSAB.35.001397} {\bibfield  {journal} {\bibinfo
  {journal} {J. Opt. Soc. Am. B}\ }\textbf {\bibinfo {volume} {35}},\ \bibinfo
  {pages} {1397} (\bibinfo {year} {2018})}\BibitemShut {NoStop}%
\bibitem [{\citenamefont {Gaponenko}(2010)}]{gaponenko_2010}%
  \BibitemOpen
  \bibfield  {author} {\bibinfo {author} {\bibfnamefont {S.~V.}\ \bibnamefont
  {Gaponenko}},\ }\href {https://doi.org/10.1017/CBO9780511750502} {\emph
  {\bibinfo {title} {Introduction to Nanophotonics}}}\ (\bibinfo  {publisher}
  {Cambridge University Press},\ \bibinfo {year} {2010})\BibitemShut {NoStop}%
\bibitem [{\citenamefont {Kristensen}\ \emph {et~al.}(2014)\citenamefont
  {Kristensen}, \citenamefont {de~Lasson},\ and\ \citenamefont
  {Gregersen}}]{kristensen_calculation_2014}%
  \BibitemOpen
  \bibfield  {author} {\bibinfo {author} {\bibfnamefont {P.~T.}\ \bibnamefont
  {Kristensen}}, \bibinfo {author} {\bibfnamefont {J.~R.}\ \bibnamefont
  {de~Lasson}},\ and\ \bibinfo {author} {\bibfnamefont {N.}~\bibnamefont
  {Gregersen}},\ }\bibfield  {title} {\bibinfo {title} {Calculation,
  normalization, and perturbation of quasinormal modes in coupled
  cavity-waveguide systems},\ }\href {https://doi.org/10.1364/OL.39.006359}
  {\bibfield  {journal} {\bibinfo  {journal} {Opt. Lett.}\ }\textbf {\bibinfo
  {volume} {39}},\ \bibinfo {pages} {6359} (\bibinfo {year}
  {2014})}\BibitemShut {NoStop}%
\bibitem [{\citenamefont {Franke}\ \emph {et~al.}(2021)\citenamefont {Franke},
  \citenamefont {Ren}, \citenamefont {Richter}, \citenamefont {Knorr},\ and\
  \citenamefont {Hughes}}]{franke_fermis_2021}%
  \BibitemOpen
  \bibfield  {author} {\bibinfo {author} {\bibfnamefont {S.}~\bibnamefont
  {Franke}}, \bibinfo {author} {\bibfnamefont {J.}~\bibnamefont {Ren}},
  \bibinfo {author} {\bibfnamefont {M.}~\bibnamefont {Richter}}, \bibinfo
  {author} {\bibfnamefont {A.}~\bibnamefont {Knorr}},\ and\ \bibinfo {author}
  {\bibfnamefont {S.}~\bibnamefont {Hughes}},\ }\bibfield  {title} {\bibinfo
  {title} {Fermi's golden rule for spontaneous emission in absorptive and
  amplifying media},\ }\href {https://doi.org/10.1103/PhysRevLett.127.013602}
  {\bibfield  {journal} {\bibinfo  {journal} {Phys. Rev. Lett.}\ }\textbf
  {\bibinfo {volume} {127}},\ \bibinfo {pages} {013602} (\bibinfo {year}
  {2021})}\BibitemShut {NoStop}%
\bibitem [{\citenamefont {Ren}\ \emph {et~al.}(2023)\citenamefont {Ren},
  \citenamefont {Franke}, \citenamefont {VanDrunen},\ and\ \citenamefont
  {Hughes}}]{ren_classical_2023}%
  \BibitemOpen
  \bibfield  {author} {\bibinfo {author} {\bibfnamefont {J.}~\bibnamefont
  {Ren}}, \bibinfo {author} {\bibfnamefont {S.}~\bibnamefont {Franke}},
  \bibinfo {author} {\bibfnamefont {B.}~\bibnamefont {VanDrunen}},\ and\
  \bibinfo {author} {\bibfnamefont {S.}~\bibnamefont {Hughes}},\ }\href
  {http://arxiv.org/abs/2305.12049} {{\selectlanguage {en}\bibinfo {title}
  {Classical {Purcell} factors and spontaneous emission decay rates in a linear
  gain medium}}} (\bibinfo {year} {2023}),\ \bibinfo {note} {arXiv:2305.12049
  [physics, physics:quant-ph]}\BibitemShut {NoStop}%
\bibitem [{\citenamefont {Leung}\ \emph {et~al.}(1994)\citenamefont {Leung},
  \citenamefont {Liu},\ and\ \citenamefont {Young}}]{leung_completeness_1994}%
  \BibitemOpen
  \bibfield  {author} {\bibinfo {author} {\bibfnamefont {P.~T.}\ \bibnamefont
  {Leung}}, \bibinfo {author} {\bibfnamefont {S.~Y.}\ \bibnamefont {Liu}},\
  and\ \bibinfo {author} {\bibfnamefont {K.}~\bibnamefont {Young}},\ }\bibfield
   {title} {{\selectlanguage {en}\bibinfo {title} {Completeness and
  orthogonality of quasinormal modes in leaky optical cavities}},\ }\href
  {https://doi.org/10.1103/PhysRevA.49.3057} {\bibfield  {journal} {\bibinfo
  {journal} {Phys. Rev. A}\ }\textbf {\bibinfo {volume} {49}},\ \bibinfo
  {pages} {3057} (\bibinfo {year} {1994})}\BibitemShut {NoStop}%
\bibitem [{\citenamefont {Anger}\ \emph {et~al.}(2006)\citenamefont {Anger},
  \citenamefont {Bharadwaj},\ and\ \citenamefont {Novotny}}]{Anger2006}%
  \BibitemOpen
  \bibfield  {author} {\bibinfo {author} {\bibfnamefont {P.}~\bibnamefont
  {Anger}}, \bibinfo {author} {\bibfnamefont {P.}~\bibnamefont {Bharadwaj}},\
  and\ \bibinfo {author} {\bibfnamefont {L.}~\bibnamefont {Novotny}},\
  }\bibfield  {title} {\bibinfo {title} {Enhancement and quenching of
  single-molecule fluorescence},\ }\href
  {https://doi.org/10.1103/PhysRevLett.96.113002} {\bibfield  {journal}
  {\bibinfo  {journal} {Phys. Rev. Lett.}\ }\textbf {\bibinfo {volume} {96}},\
  \bibinfo {pages} {113002} (\bibinfo {year} {2006})}\BibitemShut {NoStop}%
\bibitem [{\citenamefont {Noginov}\ \emph {et~al.}(2008)\citenamefont
  {Noginov}, \citenamefont {Podolskiy}, \citenamefont {Zhu}, \citenamefont
  {Mayy}, \citenamefont {Bahoura}, \citenamefont {Adegoke}, \citenamefont
  {Ritzo},\ and\ \citenamefont {Reynolds}}]{Noginov:08}%
  \BibitemOpen
  \bibfield  {author} {\bibinfo {author} {\bibfnamefont {M.~A.}\ \bibnamefont
  {Noginov}}, \bibinfo {author} {\bibfnamefont {V.~A.}\ \bibnamefont
  {Podolskiy}}, \bibinfo {author} {\bibfnamefont {G.}~\bibnamefont {Zhu}},
  \bibinfo {author} {\bibfnamefont {M.}~\bibnamefont {Mayy}}, \bibinfo {author}
  {\bibfnamefont {M.}~\bibnamefont {Bahoura}}, \bibinfo {author} {\bibfnamefont
  {J.~A.}\ \bibnamefont {Adegoke}}, \bibinfo {author} {\bibfnamefont {B.~A.}\
  \bibnamefont {Ritzo}},\ and\ \bibinfo {author} {\bibfnamefont
  {K.}~\bibnamefont {Reynolds}},\ }\bibfield  {title} {\bibinfo {title}
  {Compensation of loss in propagating surface plasmon polariton by gain in
  adjacent dielectric medium},\ }\href {https://doi.org/10.1364/OE.16.001385}
  {\bibfield  {journal} {\bibinfo  {journal} {Opt. Express}\ }\textbf {\bibinfo
  {volume} {16}},\ \bibinfo {pages} {1385} (\bibinfo {year}
  {2008})}\BibitemShut {NoStop}%
\bibitem [{\citenamefont {Corzine}\ \emph {et~al.}(1993)\citenamefont
  {Corzine}, \citenamefont {Yan},\ and\ \citenamefont
  {Coldren}}]{CORZINE199317}%
  \BibitemOpen
  \bibfield  {author} {\bibinfo {author} {\bibfnamefont {S.~W.}\ \bibnamefont
  {Corzine}}, \bibinfo {author} {\bibfnamefont {R.-H.}\ \bibnamefont {Yan}},\
  and\ \bibinfo {author} {\bibfnamefont {L.~A.}\ \bibnamefont {Coldren}},\ }in\
  \href {https://doi.org/https://doi.org/10.1016/B978-0-08-051558-8.50007-5}
  {\emph {\bibinfo {booktitle} {Quantum Well Lasers}}},\ \bibinfo {editor}
  {edited by\ \bibinfo {editor} {\bibfnamefont {P.~S.}\ \bibnamefont {Zory}}}\
  (\bibinfo  {publisher} {Academic Press},\ \bibinfo {address} {San Diego},\
  \bibinfo {year} {1993})\BibitemShut {NoStop}%
\bibitem [{SM()}]{SM}%
  \BibitemOpen
  \href@noop {} {\bibinfo {title} {See supplemental material at [url will be
  inserted by publisher] for further information on the effective mode volume,
  calculations in a dispersive gain media, and the radiative beta
  factor}}\BibitemShut {NoStop}%
\bibitem [{\citenamefont {Ge}\ \emph {et~al.}(2011)\citenamefont {Ge},
  \citenamefont {Chong}, \citenamefont {Rotter}, \citenamefont {T\"ureci},\
  and\ \citenamefont {Stone}}]{PhysRevA.84.023820_Ge}%
  \BibitemOpen
  \bibfield  {author} {\bibinfo {author} {\bibfnamefont {L.}~\bibnamefont
  {Ge}}, \bibinfo {author} {\bibfnamefont {Y.~D.}\ \bibnamefont {Chong}},
  \bibinfo {author} {\bibfnamefont {S.}~\bibnamefont {Rotter}}, \bibinfo
  {author} {\bibfnamefont {H.~E.}\ \bibnamefont {T\"ureci}},\ and\ \bibinfo
  {author} {\bibfnamefont {A.~D.}\ \bibnamefont {Stone}},\ }\bibfield  {title}
  {\bibinfo {title} {Unconventional modes in lasers with spatially varying gain
  and loss},\ }\href {https://doi.org/10.1103/PhysRevA.84.023820} {\bibfield
  {journal} {\bibinfo  {journal} {Phys. Rev. A}\ }\textbf {\bibinfo {volume}
  {84}},\ \bibinfo {pages} {023820} (\bibinfo {year} {2011})}\BibitemShut
  {NoStop}%
\bibitem [{\citenamefont {Doronin}\ \emph {et~al.}(2022)\citenamefont
  {Doronin}, \citenamefont {Andrianov},\ and\ \citenamefont
  {Zyablovsky}}]{PhysRevLett.129.133901_Doronin}%
  \BibitemOpen
  \bibfield  {author} {\bibinfo {author} {\bibfnamefont {I.~V.}\ \bibnamefont
  {Doronin}}, \bibinfo {author} {\bibfnamefont {E.~S.}\ \bibnamefont
  {Andrianov}},\ and\ \bibinfo {author} {\bibfnamefont {A.~A.}\ \bibnamefont
  {Zyablovsky}},\ }\bibfield  {title} {\bibinfo {title} {Overcoming the
  diffraction limit on the size of dielectric resonators using an amplifying
  medium},\ }\href {https://doi.org/10.1103/PhysRevLett.129.133901} {\bibfield
  {journal} {\bibinfo  {journal} {Phys. Rev. Lett.}\ }\textbf {\bibinfo
  {volume} {129}},\ \bibinfo {pages} {133901} (\bibinfo {year}
  {2022})}\BibitemShut {NoStop}%
\bibitem [{\citenamefont {Pick}\ \emph {et~al.}(2017)\citenamefont {Pick},
  \citenamefont {Zhen}, \citenamefont {Miller}, \citenamefont {Hsu},
  \citenamefont {Hernandez}, \citenamefont {Rodriguez}, \citenamefont
  {Solja{\v{c}}i{\'{c}}},\ and\ \citenamefont {Johnson}}]{pick_general_2017}%
  \BibitemOpen
  \bibfield  {author} {\bibinfo {author} {\bibfnamefont {A.}~\bibnamefont
  {Pick}}, \bibinfo {author} {\bibfnamefont {B.}~\bibnamefont {Zhen}}, \bibinfo
  {author} {\bibfnamefont {O.~D.}\ \bibnamefont {Miller}}, \bibinfo {author}
  {\bibfnamefont {C.~W.}\ \bibnamefont {Hsu}}, \bibinfo {author} {\bibfnamefont
  {F.}~\bibnamefont {Hernandez}}, \bibinfo {author} {\bibfnamefont {A.~W.}\
  \bibnamefont {Rodriguez}}, \bibinfo {author} {\bibfnamefont {M.}~\bibnamefont
  {Solja{\v{c}}i{\'{c}}}},\ and\ \bibinfo {author} {\bibfnamefont {S.~G.}\
  \bibnamefont {Johnson}},\ }\bibfield  {title} {\bibinfo {title} {General
  theory of spontaneous emission near exceptional points},\ }\href
  {https://doi.org/10.1364/oe.25.012325} {\bibfield  {journal} {\bibinfo
  {journal} {Opt. Express}\ }\textbf {\bibinfo {volume} {25}},\ \bibinfo
  {pages} {12325} (\bibinfo {year} {2017})}\BibitemShut {NoStop}%
\bibitem [{\citenamefont {Miri}\ and\ \citenamefont
  {Al\`{u}}(2019)}]{miri_exceptional_2019}%
  \BibitemOpen
  \bibfield  {author} {\bibinfo {author} {\bibfnamefont {M.-A.}\ \bibnamefont
  {Miri}}\ and\ \bibinfo {author} {\bibfnamefont {A.}~\bibnamefont {Al\`{u}}},\
  }\bibfield  {title} {\bibinfo {title} {Exceptional points in optics and
  photonics},\ }\href {https://doi.org/10.1126/science.aar7709} {\bibfield
  {journal} {\bibinfo  {journal} {Science}\ }\textbf {\bibinfo {volume}
  {363}},\ \bibinfo {pages} {eaar7709} (\bibinfo {year} {2019})}\BibitemShut
  {NoStop}%
\bibitem [{\citenamefont {Minkov}\ \emph {et~al.}(2017)\citenamefont {Minkov},
  \citenamefont {Savona},\ and\ \citenamefont {Gerace}}]{10.1063/1.4991416}%
  \BibitemOpen
  \bibfield  {author} {\bibinfo {author} {\bibfnamefont {M.}~\bibnamefont
  {Minkov}}, \bibinfo {author} {\bibfnamefont {V.}~\bibnamefont {Savona}},\
  and\ \bibinfo {author} {\bibfnamefont {D.}~\bibnamefont {Gerace}},\
  }\bibfield  {title} {\bibinfo {title} {{Photonic crystal slab cavity
  simultaneously optimized for ultra-high Q/V and vertical radiation
  coupling}},\ }\href {https://doi.org/10.1063/1.4991416} {\bibfield  {journal}
  {\bibinfo  {journal} {Applied Physics Letters}\ }\textbf {\bibinfo {volume}
  {111}},\ \bibinfo {pages} {131104} (\bibinfo {year} {2017})}\BibitemShut
  {NoStop}%
\bibitem [{\citenamefont {Vasco}\ and\ \citenamefont
  {Savona}(2021)}]{vasco2021global}%
  \BibitemOpen
  \bibfield  {author} {\bibinfo {author} {\bibfnamefont {J.}~\bibnamefont
  {Vasco}}\ and\ \bibinfo {author} {\bibfnamefont {V.}~\bibnamefont {Savona}},\
  }\bibfield  {title} {\bibinfo {title} {Global optimization of an encapsulated
  si/sio 2 l3 cavity with a 43 million quality factor},\ }\href@noop {}
  {\bibfield  {journal} {\bibinfo  {journal} {Scientific Reports}\ }\textbf
  {\bibinfo {volume} {11}},\ \bibinfo {pages} {10121} (\bibinfo {year}
  {2021})}\BibitemShut {NoStop}%
\bibitem [{\citenamefont {Granchi}\ \emph {et~al.}(2023)\citenamefont
  {Granchi}, \citenamefont {Intonti}, \citenamefont {Florescu}, \citenamefont
  {Garc{\'\i}a}, \citenamefont {Gurioli},\ and\ \citenamefont
  {Arregui}}]{granchi2023q}%
  \BibitemOpen
  \bibfield  {author} {\bibinfo {author} {\bibfnamefont {N.}~\bibnamefont
  {Granchi}}, \bibinfo {author} {\bibfnamefont {F.}~\bibnamefont {Intonti}},
  \bibinfo {author} {\bibfnamefont {M.}~\bibnamefont {Florescu}}, \bibinfo
  {author} {\bibfnamefont {P.~D.}\ \bibnamefont {Garc{\'\i}a}}, \bibinfo
  {author} {\bibfnamefont {M.}~\bibnamefont {Gurioli}},\ and\ \bibinfo {author}
  {\bibfnamefont {G.}~\bibnamefont {Arregui}},\ }\bibfield  {title} {\bibinfo
  {title} {Q-factor optimization of modes in ordered and disordered photonic
  systems using non-hermitian perturbation theory},\ }\href@noop {} {\bibfield
  {journal} {\bibinfo  {journal} {ACS photonics}\ }\textbf {\bibinfo {volume}
  {10}},\ \bibinfo {pages} {2808} (\bibinfo {year} {2023})}\BibitemShut
  {NoStop}%
\bibitem [{\citenamefont {Albrechtsen}\ \emph {et~al.}(2022)\citenamefont
  {Albrechtsen}, \citenamefont {Vosoughi~Lahijani}, \citenamefont
  {Christiansen}, \citenamefont {Nguyen}, \citenamefont {Casses}, \citenamefont
  {Hansen}, \citenamefont {Stenger}, \citenamefont {Sigmund}, \citenamefont
  {Jansen}, \citenamefont {M{\o}rk} \emph {et~al.}}]{albrechtsen2022nanometer}%
  \BibitemOpen
  \bibfield  {author} {\bibinfo {author} {\bibfnamefont {M.}~\bibnamefont
  {Albrechtsen}}, \bibinfo {author} {\bibfnamefont {B.}~\bibnamefont
  {Vosoughi~Lahijani}}, \bibinfo {author} {\bibfnamefont {R.~E.}\ \bibnamefont
  {Christiansen}}, \bibinfo {author} {\bibfnamefont {V.~T.~H.}\ \bibnamefont
  {Nguyen}}, \bibinfo {author} {\bibfnamefont {L.~N.}\ \bibnamefont {Casses}},
  \bibinfo {author} {\bibfnamefont {S.~E.}\ \bibnamefont {Hansen}}, \bibinfo
  {author} {\bibfnamefont {N.}~\bibnamefont {Stenger}}, \bibinfo {author}
  {\bibfnamefont {O.}~\bibnamefont {Sigmund}}, \bibinfo {author} {\bibfnamefont
  {H.}~\bibnamefont {Jansen}}, \bibinfo {author} {\bibfnamefont
  {J.}~\bibnamefont {M{\o}rk}}, \emph {et~al.},\ }\bibfield  {title} {\bibinfo
  {title} {Nanometer-scale photon confinement in topology-optimized dielectric
  cavities},\ }\href@noop {} {\bibfield  {journal} {\bibinfo  {journal} {Nature
  Communications}\ }\textbf {\bibinfo {volume} {13}},\ \bibinfo {pages} {6281}
  (\bibinfo {year} {2022})}\BibitemShut {NoStop}%
\bibitem [{\citenamefont {Carlson}\ and\ \citenamefont
  {Hughes}(2020)}]{PhysRevB_Chelsea}%
  \BibitemOpen
  \bibfield  {author} {\bibinfo {author} {\bibfnamefont {C.}~\bibnamefont
  {Carlson}}\ and\ \bibinfo {author} {\bibfnamefont {S.}~\bibnamefont
  {Hughes}},\ }\bibfield  {title} {\bibinfo {title} {Dissipative modes,
  {Purcell factors}, and directional beta factors in gold bowtie nanoantenna
  structures},\ }\href {https://doi.org/10.1103/PhysRevB.102.155301} {\bibfield
   {journal} {\bibinfo  {journal} {Phys. Rev. B}\ }\textbf {\bibinfo {volume}
  {102}},\ \bibinfo {pages} {155301} (\bibinfo {year} {2020})}\BibitemShut
  {NoStop}%
\bibitem [{\citenamefont {Kewes}\ \emph {et~al.}(2017)\citenamefont {Kewes},
  \citenamefont {Herrmann}, \citenamefont {Rodr\'{\i}guez-Oliveros},
  \citenamefont {Kuhlicke}, \citenamefont {Benson},\ and\ \citenamefont
  {Busch}}]{PhysRevLett.118.237402}%
  \BibitemOpen
  \bibfield  {author} {\bibinfo {author} {\bibfnamefont {G.}~\bibnamefont
  {Kewes}}, \bibinfo {author} {\bibfnamefont {K.}~\bibnamefont {Herrmann}},
  \bibinfo {author} {\bibfnamefont {R.}~\bibnamefont
  {Rodr\'{\i}guez-Oliveros}}, \bibinfo {author} {\bibfnamefont
  {A.}~\bibnamefont {Kuhlicke}}, \bibinfo {author} {\bibfnamefont
  {O.}~\bibnamefont {Benson}},\ and\ \bibinfo {author} {\bibfnamefont
  {K.}~\bibnamefont {Busch}},\ }\bibfield  {title} {\bibinfo {title}
  {Limitations of particle-based spasers},\ }\href
  {https://doi.org/10.1103/PhysRevLett.118.237402} {\bibfield  {journal}
  {\bibinfo  {journal} {Phys. Rev. Lett.}\ }\textbf {\bibinfo {volume} {118}},\
  \bibinfo {pages} {237402} (\bibinfo {year} {2017})}\BibitemShut {NoStop}%
\bibitem [{\citenamefont {Mikhailova}\ \emph {et~al.}(2019)\citenamefont
  {Mikhailova}, \citenamefont {Shaposhnikov}, \citenamefont {Tomilin},\ and\
  \citenamefont {Alentiev}}]{mikhailova2019nanostructures}%
  \BibitemOpen
  \bibfield  {author} {\bibinfo {author} {\bibfnamefont {T.}~\bibnamefont
  {Mikhailova}}, \bibinfo {author} {\bibfnamefont {A.}~\bibnamefont
  {Shaposhnikov}}, \bibinfo {author} {\bibfnamefont {S.}~\bibnamefont
  {Tomilin}},\ and\ \bibinfo {author} {\bibfnamefont {D.}~\bibnamefont
  {Alentiev}},\ }\bibfield  {title} {\bibinfo {title} {Nanostructures with
  magnetooptical and plasmonic response for optical sensors and nanophotonic
  devices},\ }in\ \href@noop {} {\emph {\bibinfo {booktitle} {Journal of
  Physics: Conference Series}}},\ Vol.\ \bibinfo {volume} {1410}\ (\bibinfo
  {organization} {IOP Publishing},\ \bibinfo {year} {2019})\ p.\ \bibinfo
  {pages} {012163}\BibitemShut {NoStop}%
\bibitem [{\citenamefont {Gao}\ \emph {et~al.}(2020)\citenamefont {Gao},
  \citenamefont {Wu}, \citenamefont {Zhou}, \citenamefont {Yang}, \citenamefont
  {Guo}, \citenamefont {Wang}, \citenamefont {You}, \citenamefont {Gu},
  \citenamefont {Lu}, \citenamefont {Gong},\ and\ \citenamefont
  {Tong}}]{gao_nanolaser_2020}%
  \BibitemOpen
  \bibfield  {author} {\bibinfo {author} {\bibfnamefont {Y.}~\bibnamefont
  {Gao}}, \bibinfo {author} {\bibfnamefont {H.}~\bibnamefont {Wu}}, \bibinfo
  {author} {\bibfnamefont {N.}~\bibnamefont {Zhou}}, \bibinfo {author}
  {\bibfnamefont {Y.}~\bibnamefont {Yang}}, \bibinfo {author} {\bibfnamefont
  {X.}~\bibnamefont {Guo}}, \bibinfo {author} {\bibfnamefont {P.}~\bibnamefont
  {Wang}}, \bibinfo {author} {\bibfnamefont {J.}~\bibnamefont {You}}, \bibinfo
  {author} {\bibfnamefont {Y.}~\bibnamefont {Gu}}, \bibinfo {author}
  {\bibfnamefont {G.}~\bibnamefont {Lu}}, \bibinfo {author} {\bibfnamefont
  {Q.}~\bibnamefont {Gong}},\ and\ \bibinfo {author} {\bibfnamefont
  {L.}~\bibnamefont {Tong}},\ }\bibfield  {title} {\bibinfo {title}
  {Single-nanorod plasmon nanolaser: A route toward a three-dimensional
  ultraconfined lasing mode},\ }\href
  {https://doi.org/10.1103/PhysRevA.102.063520} {\bibfield  {journal} {\bibinfo
   {journal} {Phys. Rev. A}\ }\textbf {\bibinfo {volume} {102}},\ \bibinfo
  {pages} {063520} (\bibinfo {year} {2020})}\BibitemShut {NoStop}%
\bibitem [{\citenamefont {Caicedo}\ \emph {et~al.}(2022)\citenamefont
  {Caicedo}, \citenamefont {Cathey}, \citenamefont {Infusino}, \citenamefont
  {Aradian},\ and\ \citenamefont {Veltri}}]{Caicedo:22}%
  \BibitemOpen
  \bibfield  {author} {\bibinfo {author} {\bibfnamefont {K.}~\bibnamefont
  {Caicedo}}, \bibinfo {author} {\bibfnamefont {A.}~\bibnamefont {Cathey}},
  \bibinfo {author} {\bibfnamefont {M.}~\bibnamefont {Infusino}}, \bibinfo
  {author} {\bibfnamefont {A.}~\bibnamefont {Aradian}},\ and\ \bibinfo {author}
  {\bibfnamefont {A.}~\bibnamefont {Veltri}},\ }\bibfield  {title} {\bibinfo
  {title} {Gain-driven singular resonances in active core-shell and nano-shell
  plasmonic particles},\ }\href {https://doi.org/10.1364/JOSAB.441637}
  {\bibfield  {journal} {\bibinfo  {journal} {J. Opt. Soc. Am. B}\ }\textbf
  {\bibinfo {volume} {39}},\ \bibinfo {pages} {107} (\bibinfo {year}
  {2022})}\BibitemShut {NoStop}%
\bibitem [{\citenamefont {Kockum}\ \emph {et~al.}(2019)\citenamefont {Kockum},
  \citenamefont {Miranowicz}, \citenamefont {Liberato}, \citenamefont
  {Savasta},\ and\ \citenamefont {Nori}}]{FriskKockum2019}%
  \BibitemOpen
  \bibfield  {author} {\bibinfo {author} {\bibfnamefont {A.~F.}\ \bibnamefont
  {Kockum}}, \bibinfo {author} {\bibfnamefont {A.}~\bibnamefont {Miranowicz}},
  \bibinfo {author} {\bibfnamefont {S.~D.}\ \bibnamefont {Liberato}}, \bibinfo
  {author} {\bibfnamefont {S.}~\bibnamefont {Savasta}},\ and\ \bibinfo {author}
  {\bibfnamefont {F.}~\bibnamefont {Nori}},\ }\bibfield  {title} {\bibinfo
  {title} {Ultrastrong coupling between light and matter},\ }\href
  {https://doi.org/10.1038/s42254-018-0006-2} {\bibfield  {journal} {\bibinfo
  {journal} {Nature Reviews Physics}\ }\textbf {\bibinfo {volume} {1}},\
  \bibinfo {pages} {19} (\bibinfo {year} {2019})}\BibitemShut {NoStop}%
\bibitem [{\citenamefont {Forn-D\'{\i}az}\ \emph {et~al.}(2019)\citenamefont
  {Forn-D\'{\i}az}, \citenamefont {Lamata}, \citenamefont {Rico}, \citenamefont
  {Kono},\ and\ \citenamefont {Solano}}]{RevModPhys.91.025005}%
  \BibitemOpen
  \bibfield  {author} {\bibinfo {author} {\bibfnamefont {P.}~\bibnamefont
  {Forn-D\'{\i}az}}, \bibinfo {author} {\bibfnamefont {L.}~\bibnamefont
  {Lamata}}, \bibinfo {author} {\bibfnamefont {E.}~\bibnamefont {Rico}},
  \bibinfo {author} {\bibfnamefont {J.}~\bibnamefont {Kono}},\ and\ \bibinfo
  {author} {\bibfnamefont {E.}~\bibnamefont {Solano}},\ }\bibfield  {title}
  {\bibinfo {title} {Ultrastrong coupling regimes of light-matter
  interaction},\ }\href {https://doi.org/10.1103/RevModPhys.91.025005}
  {\bibfield  {journal} {\bibinfo  {journal} {Rev. Mod. Phys.}\ }\textbf
  {\bibinfo {volume} {91}},\ \bibinfo {pages} {025005} (\bibinfo {year}
  {2019})}\BibitemShut {NoStop}%
\bibitem [{\citenamefont {Salmon}\ \emph {et~al.}(2022)\citenamefont {Salmon},
  \citenamefont {Gustin}, \citenamefont {Settineri}, \citenamefont {Stefano},
  \citenamefont {Zueco}, \citenamefont {Savasta}, \citenamefont {Nori},\ and\
  \citenamefont {Hughes}}]{Salmon2022}%
  \BibitemOpen
  \bibfield  {author} {\bibinfo {author} {\bibfnamefont {W.}~\bibnamefont
  {Salmon}}, \bibinfo {author} {\bibfnamefont {C.}~\bibnamefont {Gustin}},
  \bibinfo {author} {\bibfnamefont {A.}~\bibnamefont {Settineri}}, \bibinfo
  {author} {\bibfnamefont {O.~D.}\ \bibnamefont {Stefano}}, \bibinfo {author}
  {\bibfnamefont {D.}~\bibnamefont {Zueco}}, \bibinfo {author} {\bibfnamefont
  {S.}~\bibnamefont {Savasta}}, \bibinfo {author} {\bibfnamefont
  {F.}~\bibnamefont {Nori}},\ and\ \bibinfo {author} {\bibfnamefont
  {S.}~\bibnamefont {Hughes}},\ }\bibfield  {title} {\bibinfo {title}
  {Gauge-independent emission spectra and quantum correlations in the
  ultrastrong coupling regime of open system cavity-{QED}},\ }\href
  {https://doi.org/10.1515/nanoph-2021-0718} {\bibfield  {journal} {\bibinfo
  {journal} {Nanophotonics}\ }\textbf {\bibinfo {volume} {11}},\ \bibinfo
  {pages} {1573} (\bibinfo {year} {2022})}\BibitemShut {NoStop}%
\end{thebibliography}%

\end{document}